\documentclass[traditabstract]{aa} 
\usepackage{longtable}
\usepackage{latexsym}
\usepackage{amssymb}
\usepackage{lscape}
\usepackage[]{natbib}

\usepackage{color}

\usepackage{graphicx}
\usepackage{txfonts}
\def\lsim{\mathrel{\rlap{\lower 3pt \hbox{$\sim$}} \raise 2.0pt \hbox{$<$}}}
\def\gsim{\mathrel{\rlap{\lower 3pt \hbox{$\sim$}} \raise 2.0pt \hbox{$>$}}}
\title{MUSE sneaks a peek at extreme ram-pressure events - III. Tomography of UGC 6697, a massive galaxy falling into Abell 1367}
\subtitle{}
\author{G. Consolandi \inst{1}                               
\and G. Gavazzi \inst{1}                               
\and M. Fossati \inst{2,3}                                  
\and M. Fumagalli \inst{4}                                
\and A. Boselli \inst{5}                                    
\and M. Yagi \inst{6,7}                            
\and M. Yoshida \inst{8}
}
\authorrunning{G. Consolandi et al.}
\titlerunning{Tomography of UGC 6697, a massive galaxy falling into Abell 1367}
\institute{Universit\`a degli Studi di Milano-Bicocca, Piazza della Scienza 3, 20126 Milano, Italy\\
\email {guido.consolandi@mib.infn.it}
\and
Max-Planck-Institut f{\"u}r Extraterrestrische Physik, Giessenbachstrasse, D-85748 Garching, Germany\\
\email {mfossati@mpe.mpg.de}
\and
Universit{\"a}ts-Sternwarte M{\"u}nchen, Schenierstrasse 1, D-81679 M{\"u}nchen, Germany. 
\and
Institute for Computational Cosmology, and Centre for Extragalactic Astronomy, Durham University, South Road, Durham, DH1 3LE, UK\\
\email {michele.fumagalli@durham.ac.uk}
\and
Aix Marseille Univ, CNRS, LAM, Laboratoire d'Astrophysique de Marseille, Marseille, France\\
\email {alessandro.boselli@lam.fr}
\and
Optical and Infrared Astronomy Division, National Astronomical Observatory of Japan, Mitaka, Tokyo, 181-8588, Japan\\
\email {yagi.masafumi@nao.ac.jp}
\and
Graduate School of Science and Engineering, Hosei University, 3-7-2, Kajinocho, Koganei, Tokyo, 184-8584 Japan
\and
Hiroshima Astrophysical Science Center, Hiroshima University, 1-3-1, Kagamiyama, Higashi-Hiroshima, Hiroshima, 739-8526, Japan\\
}
\begin{document}
\date{Received; accepted}
\abstract{
We present the MUSE observations of UGC 6697, a giant (M$_*\approx 10^{10}\rm M_\odot$) spiral galaxy infalling in the nearby cluster Abell 1367.
During its high velocity transit through the intracluster medium (ICM), the hydrodynamical interactions with the ICM produce a $\approx 100\rm~kpc$ tail of ionized gas
that we map with a mosaic of five MUSE pointings up to 60 kpc from the galaxy. 
CGCG 97087N, a small companion that lies at few arcminutes in projection from UGC 6697, is also suffering from the hydrodynamic action of the ICM of the cluster. 
Along the whole extent of the tail we detect diffuse H$\alpha$ emission and, to a lesser extent, H$\beta$, [OIII]$\lambda5007$, and [OI]$\lambda6300$.
By comparing the kinematics and distribution of gas and stars (as traced by the CaII triplet) for both galaxies, we separate the ionized gas, as traced by the H$\alpha$ line, 
in a component still bound to the galaxy and
a component that is stripped. We find that the ''onboard'' component shows low 
velocity dispersion and line ratios consistent with photoionization by hot stars. The stripped gas is  more turbulent, 
with velocity dispersions up to $\gtrsim \rm100 ~km~s^{-1}$, and is excited by shocks as traced by  high values of [OI]/H$\alpha$ and [NII]/H$\alpha$ ratio.
In the tail of UGC 6697 we identify numerous bright compact knots with line ratios typical of HII regions. 
These are distributed along the only streams of stripped gas retaining low velocity dispersions ($\lesssim\rm 35 ~km~s^{-1}$).  
Despite being in the stripped gas, their physical properties do not differentiate from normal HII regions in galactic disks.  
We find evidence of a past fast encounter between the two galaxies in the form of a double tail emerging from CGCG 97087N that connects with UGC 6697.
This encounter might have increased the efficiency of the
stripping process, leaving the stellar distribution and kinematics unaltered.\\ 
 } 
\keywords{Galaxies: evolution -- Galaxies:  clusters: individual: UGC 6697  -- Galaxies: interactions -- Galaxies: star formation}
\maketitle
%

\section{Introduction}
The evolutionary path that drives galaxies through the cosmic ages from being actively star forming to "red and dead" systems 
is regulated by the complex interplay between two broad classes of processes which are able to remove (or rapidly consume) the reservoir of cold gas, thus 
quenching the galaxies star formation: i) internal (or secular) processes that are driven by the mass \citep[dubbed  "downsizing";][]{fonta09, pg09, siu16}; 
ii) external processes that are driven by the environment in which galaxies are embedded.
As galaxies approach regions of high density, such as groups or clusters, environmental processes start to sum up to the secular ones 
strongly contributing to the build up of the red and passive population of galaxies \citep{dress80,bos06}.\\
Environmental processes can be further divided into two classes: gravitational interactions 
\citep[tidal interaction - harassment; ][]{harassment} and hydrodynamic interactions between the hot intracluster medium (ICM)  
and the interstellar medium (ISM) of  galaxies infalling at high velocity in clusters (ram-pressure, Gunn \& Gott 1972; viscous stripping, Nulsen 1982; thermal evaporation, Cowie \& McKee 1977;
and starvation, Larson 1980. See Boselli \& Gavazzi 2006; 2014 for reviews). 
However, the relative importance of  the different processes shaping galaxies in dense environments
is still poorly defined. 
The direct observations of environmental interactions has been often hampered by technological limitations.
All multiwavelength studies aiming at directly detect environmental processes demand a high sensitivity as well as sufficient resolution.
Apart from striking merging events, shells and stellar streams triggered by mergers are tenuous features \citep[see][]{duc15} characterized by
low surface brightness emission. Similarly, the diffuse gas stripped by hydrodynamic interactions is characterized by emission of very low surface brightness \citep{yoshi04,sun07,VIVA,1690,fuma14,fox16,merluzzi16}.\\
The local Universe is obviously the best place to reach high sensitivities with high resolution in a reasonable integration time. On the other hand local galaxies have apparent sizes 
of several arcminutes, thus requiring large fields-of-views \citep{yoshi04,1690}.
Hence, for many years, the direct observations of the environmental interactions has been limited to few targeted objects, preventing from a statistical analysis.
\begin{figure*}
\centering
\includegraphics[width=19cm]{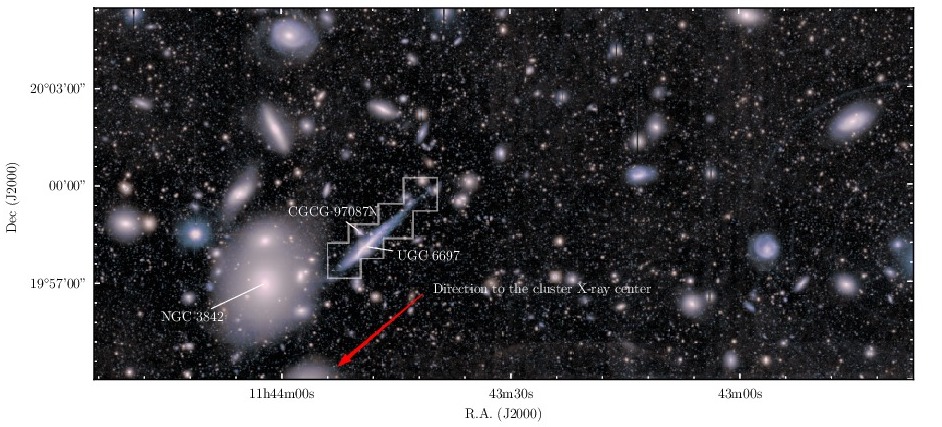}
\caption{RGB (iRB) image of  the central region of Abell 1367 containing 
 NGC 3842, one of the two giant E galaxies in this cluster (left), the highly inclined UGC 6697 and its faint companion CGCG 97087N, which are the subject of the present
 investigation. The gray line indicates the positions of the MUSE mosaic.  The observations were taken with the Subaru observatory by Yagi et al. (2017).}
\label{subaru}  
\end{figure*}
Recently however we entered a new era. 
The advent of new imaging cameras and Integral Field Units (IFUs) spectrographs  is  a turning point for the study of the effects of the environment.
In visible light, telescopes such as CFHT and SUBARU now mount cameras with large field of views (FOVs, $\approx 1 deg^2$) and are equipped with large interferometric filters, making possible
to map large areas.\\ 
For instance, using SuprimeCam mounted on the SUBARU telescope, \citet{yagi10} and \citet{yagi17} have mapped the ionized gas emission as traced 
by H$\alpha$ of galaxies in the Coma and Abell 1367 clusters with unprecedented spatial resolution and sensitivity. They unveiled that stripping phenomena are
ubiquitous in these clusters as a large fraction of late-type galaxies (LTGs) are observed in the process of loosing their ionized gas because of their hydrodynamic interaction with the ICM. 
More recently, \citet{1690}, using the MegaCam mounted on the CFHT telescope, targeted with deep H$\alpha$ observations a massive spiral 
in the Virgo cluster (NGC 4569) and unveiled a spectacular 75 kpc tail trailing behind the galaxy. 
These observations are particularly interesting because vividly show that ram pressure stripping may be very effective even in massive galaxies inside a young and intermediate-mass cluster, such as Virgo
\citep{bing87}.\\ 
In this "new era", 
IFUs represent a breaktrough in the study of environment.
As shown by the recent studies of \citet{fuma14}, \citet{fox16}, \citet{merluzzi16}, \citet{gasp1}, \citet{gasp2}, and \citet{gasp3}, IFUs are able to recover with high spatial resolution
the dynamics of both gas and stars and to provide all the spectroscopic informations to constrain the physical state of the ISM and of the stellar component.
Many IFU follow-up observations  of galaxies interacting in high density regions are already scheduled \citep{gasp1}, 
allowing us to build sufficient statistics to explore characteristic trends in the 
physical state of systems interacting with the environment.\\
The present work analyzes the MUSE observations of the spiral galaxy UGC 6697 infalling in the nearby cluster Abell 1367.
Unlike the galaxy ESO 137-001 \citep[the very first studied with this instrument,][]{fuma14,fox16}, UGC 6697 is a massive galaxy (M$_*\approx10^{10}$M$_\odot$) transiting 
edge-on through the ICM of the cluster.
Such configuration should, in principle, disfavour the ram-pressure stripping.  
A small companion, CGCG 97087N, lies at just few arcseconds in projection from UGC 6697 and recent SUBARU observation
reveals double gaseous tails connecting to UGC 6697 and hinting at a possible galaxy-galaxy intraction. Such system, differently from ESO 137-001, 
offers the opportunity to study a triple (ICM-galaxy-satellite) interaction. \\

The layout of the paper is as follows.
In section \ref{sec_targets} we present the targets (UGC 6697 and CGCG 97087N) by summarizing results from the literature.
In sections \ref{Obs} and \ref{datared} we provide the details of our observations, of the data reduction, and of the software used to analyze the final data product.
The results are presented in section \ref{sec_results}, with a full kinematical analysis in sections \ref{sec_velocities}, and \ref{sec_2vel}. 
The analysis of the physical state of the ionized gas using line ratios is presented in section \ref{sec_chemis}. We further constrain the 
ionization state of the gas building spatially resolved BPT diagrams \citep{bpt} in section \ref{sec_bpt} and we extract the properties of the HII regions observed in the tails in section \ref{sec_hii}.
Finally, in section \ref{sec_discuss} we discuss our results.
\section{Targets} 
\label{sec_targets}
UGC 6697 (aka CGCG 97087, Nilson 1973, Zwicky et al. 1968) is the second brightest spiral/Irr member ($cz=6727~\rm km ~s^{-1}$) of the nearby cluster
Abell 1367 ($\langle cz \rangle=6595 \rm ~km~s^{-1}$). It is located at R.A.=$11^{h}43^{m}49^{s}.03$ and Dec=$+19^{o}58'05".5$, 
at a projected distance of 462 kpc in the NW direction from the X-ray center of Abell 1367 \citep[RA= $11^{h}44^{m}36^{s}.5$ and dec= $+19^{o}45'32"$, according to][]{piff11}, 
well within the virial radius of the cluster (2.13 Mpc; Girardi et al. 1998). The region of the cluster inhabited by UGC 6697 is shown in the Subaru field of Fig. \ref{subaru}.
A smaller companion (CGCG 97087N) with a recessional velocity of $cz=7542 ~\rm km ~s^{-1}$ lies at$\sim 20$ arcsec in the NE direction from  UGC 6697,
while NGC 3842 ($cz=6247 ~\rm km ~s^{-1}$), one of the two giant E galaxies of Abell 1367, lies at $\sim 3.5$ arcminutes in the E direction.
The small offset between the velocity of UGC 6697 and of the cluster suggests that the galaxy is crossing Abell 1367 nearly in the plane of the sky.
On the contrary, the velocity of CGCG 97087N suggests that its motion is almost along the line of sight 
\footnote{At the assumed distance of 94.8 Mpc of Abell 1367, 1 arcsec separation corresponds to 0.46 kpc, one arcmin to 27.8 kpc.}.\\
UGC 6697 and CGCG 97087N have a stellar mass  of $\approx 10^{10}$M$_{\odot}$ and $\approx 10^{9}$M$_{\odot}$ respectively, as computed from the $i$-band 
luminosity and the ($g$-$i$) color following Zibetti et al. (2009). 

UGC 6697 was pointed to our attention owing to the head-tail morphology of the radio continuum source associated with it (Gavazzi 1978). 
This morphology, generally found in elliptical cluster galaxies, had not been found in a spiral galaxy. This feature suggests that the galaxy is under the 
action of ram-pressure (Gunn \& Gott 1972) during its high velocity, edge-on transit through the cluster ICM.
Many multifrequency observations targeting UGC 6697  have been gathered after its discovery. 
From 1984 to 1989, the galaxy has been observed in H$\alpha$ (Bothun \& Schommer 1984), radio, and optical
continuum (Gavazzi et al. 1984), at 1415 MHz (Gavazzi \& Jaffe 1987) and in HI (Gavazzi  1989).
All observations consistently show signs of the cometary shape of UGC 6697 and mutually reinforce the interpretation of this object based on the ram pressure scenario.
More recently, Sun and Vikhlinin (2005), using the Chandra observatory, detected significant X-ray diffuse emission showing a morphological 
asymmetry similar to the one observed in the radio, hence suggesting an ongoing hydrodynamic interaction with the ICM of Abell 1367.
Based on the early work of \citet{Nulsen82}, these authors note that the geometry of the impact of the galaxy with the ICM (edge-on)
should disfavor ram-pressure as the main hydrodynamic process at play. Instead viscous or turbulent stripping should be the principle hydrodynamic 
process that is stripping the gas from UGC 6697.
\begin{table*}
\caption{Details of the observing blocks composing the mosaic of UGC 6697 ordered in right ascension from E to W.}								         
\centering										          
\begin{tabular}{c c c c c c c}								          
\hline\hline										          
  OB  &    RA		 &    Dec	& Obs date   & T exp	& AM &  Seeing \\	          
      &  (hh m s)  	 & ($^o$ ' ")	& (dd mm yy)   & (sec)	&    &  (arcsec) \\	          
\hline   										          
  11  &    11 43 51.46   &   19 57 42.5 &  29 02 16  & 2x2190  & 1.48-1.57   & 0.77-1.07    \\
  1   &    11 43 48.70   &   19 58 16.8 &  10 02 16  & 2x2190  & 1.40-1.45   & 1.01-1.36    \\    
  7   &    11 43 48.70   &   19 58 16.8 &  10 02 16  & 2x2190  & 1.51-1.62   & 0.72-1.95    \\    
  2   &    11 43 45.00   &   19 58 53.0 &  10 02 16  & 2x2190  & 1.40-1.44   & 1.00-1.90    \\    
  8   &    11 43 41.85   &   19 59 40.0 &  29 02 16  & 2x2190  & 1.40-1.43   & 1.02-1.43    \\    
\hline 
\end{tabular}   									          
\label{tab1}    									          
\end{table*}    

However,  variations on the ram pressure scenario were also proposed mainly because of the existence of multiple velocity
components along the line of sight in the  rotation curve of UGC 6697 (Gavazzi et al. 1984; Gavazzi et al. 2001).
Gavazzi et al. (1984) speculated that a possible collision between UGC 6697 and CGCG 97087N might have occurred, 
causing a flow of gas to emerge from UGC 6697, while Gavazzi et al. (2001) speculated that the complex morphology and kinematics 
of the galaxy was the result of the merging of two different systems.
Recently, Yagi et al. (2017), using the 8m Subaru telescope, obtained a deep H$\alpha$ image of Abell 1367, including UGC 6697.
They discovered that its low surface brightness H$\alpha$ cometary trail extends towards the NW for up to 100 kpc.
Surprisingly, they also found clear evidence of a faint gaseous double-tail that emerges perpendicularly from the disk of
CGCG 97087N and connects to the disk of UGC 6697, supporting the scenario proposed by Gavazzi et al. (1984).
\section{Observations}
\label{Obs}
\begin{figure*}[]
\centering
\includegraphics[width=19.5cm, trim = 0cm 0cm 0cm 0cm]{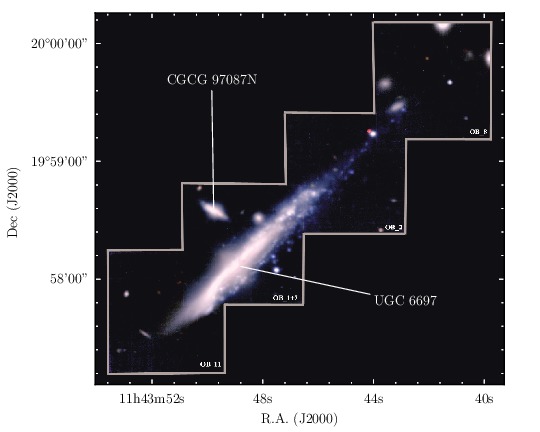}
\caption{False color image of the 5 mosaicked MUSE pointings that cover UGC 6697. 
Its companion galaxy UGC 6697N lies at $\sim20$ arcsec projected distance in the NE direction. Each RGB channel was constructed from the reduced MUSE data cube by 
projecting the cube respectively  in the $g$, $r$ and $i$ SDSS bands. 
}
\label{cube}  
\end{figure*}

UGC 6697 and its companion CGCG 97087N were observed with MUSE on VLT (UT4) on February the $10^{th}$ and $29^{th}$ 
2016, as part of proposal 096.B-0019(A) (PI Gavazzi).
A mosaic of the five MUSE pointings  taken at the coordinates listed in Table \ref{tab1} is shown in Fig. \ref{cube}.
These pointings cover from the optical front of the galaxy (South East of the mosaic) up to 54 kpc outside the optical extent of the
galaxy in the NW direction where the H$\alpha$ tail extends.
Each observation consisted of two 21.5 minutes exposures obtained rotating the instrument by 90 degrees, interspersed by an empty sky exposure
of 4 min duration, taken at the coordinates R.A.$=11^h 43^m 41^s$; Dec$=+19^o 54' 52"$.
The central pointing that covers the center of UGC 6697 and its companion CGCG 97087N was repeated (OB1 and OB7) in order to reach a higher signal 
to noise in the area of a possible interaction between the two. 
Observations were conducted while the galaxy was transiting at airmass $\sim$ 1.5 in clear 
sky conditions and seeing ranging from 0.7 to 1.9 arcsec, using the Wide Field Mode with nominal wavelength coverage (4800-9300 $\AA$). 
Flux calibration was achieved by means of observations of the standard star GD153 exposed for 160 seconds.
\section{Data reduction and analysis}
\label{datared}
The final datacube mosaic was obtained using the MUSE data reduction pipeline (v 1.2.1) and a set of Python based codes that we designed in order to 
improve the quality of the illumination correction and of the sky subtraction for the observed field.\\ 
For each exposure composing the mosaic as well as for each sky and standard star frames, the first step was to create
a master bias, master dark, a master flat and a wavelength solution within the MUSE pipeline.
These were applied to each science and sky cubes and to the standard star (GD153). 
After computing and applying the telluric correction spectrum within the MUSE pipeline, we reconstruct the final cubes for each exposure.
In order to produce a mosaic containing 5 different pointings we construct an astrometry reference frame building a regular 2D-WCS grid that
maps all the exposures of the final mosaic. Starting from the WCS information 
contained in the reference WCS frame, we build for each exposure a regular 3-D WCS grid with a 0.2" spatial pixel scale and a 1.25 $\AA$ spectral step 
from 4750 $\AA$ to 9300 $\AA$.
All science exposures are projected onto their 3-D grid using the MUSE pipeline task {\it muse\_scipost}.
Every reconstructed cube of each exposure is mapped onto the same wavelength and spatial grid.\\
The reconstructed cubes exhibit a residual variation in the illumination of each IFU that depends on the wavelength at which the FOV is observed.
In order to perform a robust sky subtraction that minimizes the residuals \citep{fuma14, fuma17} we apply to each sky and science exposure an illumination correction. 
This was computed using in-house Python codes that evaluate, after masking bright contaminating objects, the illumination correction 
as a function of wavelength by computing  the response of each IFU compared to the average response of the field in intervals of 500 $\AA$.\\

After correcting the illumination of each sky and science fields, we obtain a master sky spectrum by collapsing onto the wavelength axis the single 
sky exposure which is free of contaminating sources and properly masked. To construct a correct model of the sky spectrum 
that accounts for both the time dependency of air glow line and continuum fluxes we use SKYCORR (Noll et al. 2014) as 
in Fossati et al. (2016). The sky model is then subtracted from each spaxel interpolating the spectrum on the wavelength axis 
with a spline function. The uncertainties 
are appropriately added to the variance extension of the datacube.

After sky subtraction, a single datacube for each pointing is created by combining the respective exposures and the final 
datacube is assembled following the mosaic map. The final result is shown in the RGB of Fig.\ref{cube} constructed by convolving
the datacube with the SDSS $g$$r$$i$ filters.   
The noise background, in the central field covered by 4 expositions, has a 1$\sigma$ limiting surface brightness around the redshifted H$\alpha$ line ($\lambda\sim6680-6750 \rm \AA$) of 
$\rm \approx 5\times 10^{-19}erg~s^{-1}~cm^{-2}~arcsec^{-2}~\AA^{-1}$.
In the other fields the sky surface brightness reaches $\rm \approx 9.5\times 10^{-19}erg~s^{-1}~cm^{-2}~arcsec^{-2}~\AA^{-1}$. Around the redshifted H$\beta$ line ($\lambda\sim4930-5015 \rm \AA$) these values become 
$\rm \approx 1\times 10^{-18}erg~s^{-1}~cm^{-2}~arcsec^{-2}~\AA^{-1}$ and $\rm \approx 2\times 10^{-18}erg~s^{-1}~cm^{-2}~arcsec^{-2}~\AA^{-1}$, respectively.
In order to test the flux calibration reached within our datacubes we compare the flux of few point-like sources in the field with their correspective flux in the SDSS, finding consistency within $\approx3\%$.

In order to characterize the properties of the ionized gaseous component of the galaxies we extract flux maps of the emission lines 
listed in Table \ref{tab2} while, 
in order to trace the kinematics of the stellar component, we fit the second line of the Ca II triplet at $8542.09 \rm \AA$.
\begin{table}
\caption{Emission lines considered in this study. Column (1) is the name of the emission lines and column
(2) the respective wavelengths in air. Column (3) represent the surface brightness limits at 5$\sigma$ from the central cube covered by four exposures and
the surface brightness limits at 5$\sigma$ for the other cubes.}								         
\centering										          
\begin{tabular}{c c c }								          
\hline\hline										          
  Line       &    $\lambda$   &   $\mu_{min}$   		           \\ 	     
             &     ($\AA$)      &    (erg s$^{-1}$ cm$^{-2}$ $\AA^{-1}$ arcsec$^{-2}$)   \\	 
\hline   			        		    
  H$\beta     $   &     4861.33    &  $\approx5.3-9.7\times 10^{-18}$	  \\
  $\rm [OIII] $   &     4658.81	   &  $\approx4.9-9.1\times 10^{-18}$	  \\
  $\rm [OIII] $   &     5006.84	   &  $\approx4.9-9.1\times 10^{-18}$	  \\
  $\rm [OI]   $   &     6300.30	   &  $\approx 4.1-5.6\times 10^{-18}$	  \\
  $\rm [NII]  $   &     6548.05    &  $\approx 3-4.6\times 10^{-18}$	  \\
  H$\alpha    $   &     6562.82    &  $\approx 3-4.6\times 10^{-18}$	  \\
  $\rm [NII]  $   &     6583.45    &  $\approx 3-4.6\times 10^{-18}$	  \\
  
\hline 
\end{tabular}   									          
\label{tab2}    									          
\end{table}    
To robustly estimate emission line fluxes, we have to account for the  stellar absorptions underlying the Balmer emission lines falling in the MUSE spectral window: 
H$\alpha$ and H$\beta$. 
For this, we use the code GANDALF \citep{gandalf} complemented by the Penalize Pixel-Fitting code \citep{ppxf} to simultaneously
model the stellar continuum and the emission lines in individual spaxels with ${\rm S/N}> 5$.
The stellar continuum is modeled with the superposition of stellar templates from the MILES library \citep{miles_lib} convolved by the stellar line-of-sight velocity distribution, whereas
the emission lines and kinematics are modeled assuming a Gaussian profile. 
In each spaxel, the modeled stellar continuum spectrum is subtracted from the observed spectrum obtaining a final datacube of pure emission lines, free of stellar absorption.
\subsection{Emission line measurements}
\label{measure}
In order to account for more complex kinematics along the line of sight \citep[already known from][]{pg01} we designed a Python fitting code called $QBfit$.
$QBfit$ allows to fit groups of emission lines (or line sets) at once, such as H$\alpha$ and $[\rm NII] ~\lambda \lambda 6548,6584\AA$, by fitting 1D Gaussian functions with relative velocity 
separation of the lines kept fixed according to the wavelengths listed in Table \ref{tab2}. The fit of [OIII] and [NII] line sets assume 
the ratio between the [OIII] lines and between the [NII] lines fixed to the values published in \citet{lineratios}.
The continuum is evaluated locally with a flat power-law and the S/N ratio of the H$\alpha$ is evaluated using the "stat" extension of the datacube.
Moreover $QBfit$  allows the fit of an additional lineset with a different velocity component by fitting simultaneously the same lineset two times with a shift in wavelength.
In the end, the code returns kinematics and flux maps separated for a low velocity and an high velocity component.

We fit the emission lines in the continuum free datacube, assuming a systemic redshift of $z_{sys}=0.02243$ for UGC 6697 and after 
a median smoothing of 10x10 spaxel (2 arcsec) in the spatial dimension to increase the S$/$N ratio per pixel.
We do not perform any smoothing along the spectral dimension.
During the fitting procedure, sky line residuals and spaxels where the S/N of the H$\alpha$ line is lower than 5 are masked.
Further masking is applied to spaxel in which the line centroid or the line width are extremely deviant or showing an error greater than 50 $\rm km~s^{-1}$.\\
\begin{figure}
\centering
\includegraphics[width=9cm]{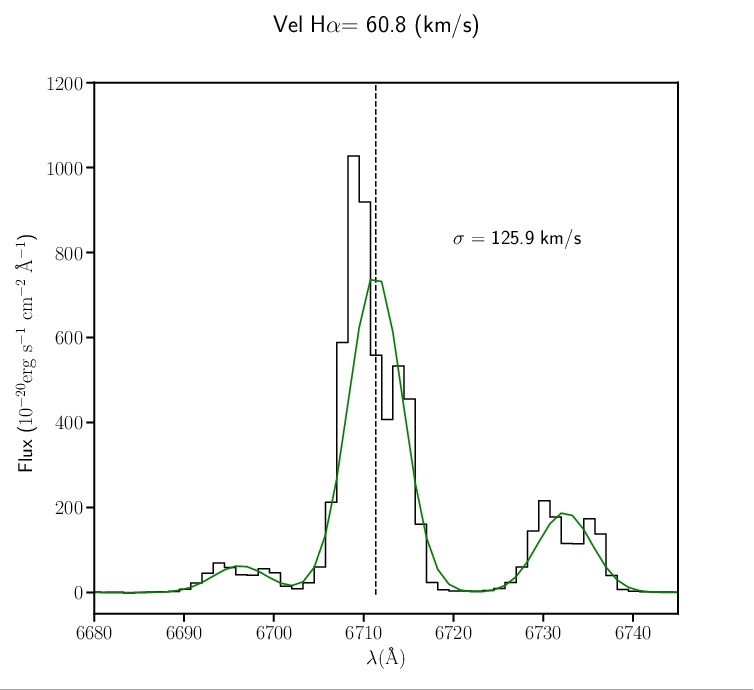}
\includegraphics[width=9cm]{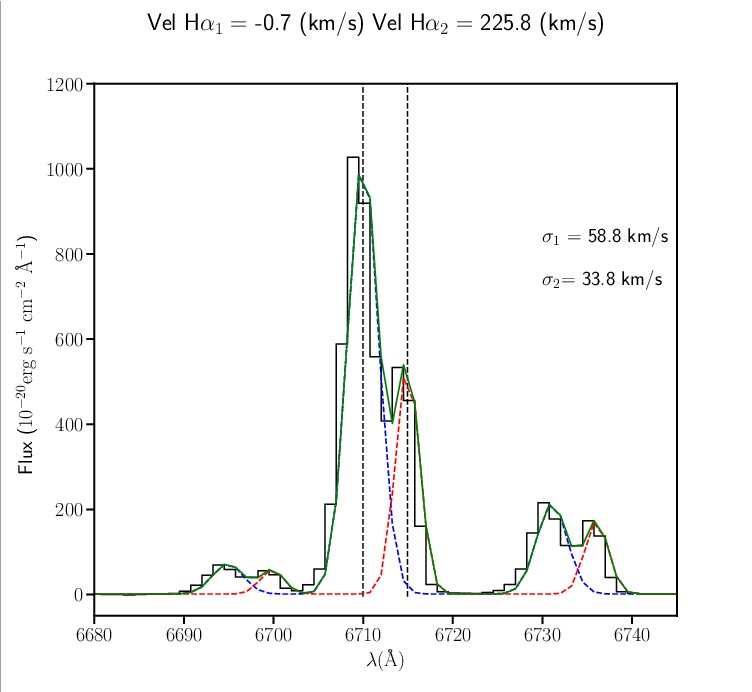}
\caption{Spectrum around H$\alpha$ of a spaxel clearly showing two velocity 
components. In the top panel, the single velocity fit to the data is shown in green, while in the bottom panel 
the sum of the two velocity component fit is shown in green. The individual velocity components are displayed with blue and red lines.
}
\label{doble}  
\end{figure}
The double velocity component fit is not necessary across the whole galaxy extension but the procedure automatically recognizes the spaxels 
that show double velocities by fitting a single velocity component and by using the width of the emission line as a proxy for multiple velocity components.
The width is evaluated after taking into account the instrumental width of the line, evaluated given the MUSE spectral resolution function.
As an example, Figure \ref{doble} shows the spectrum around the H$\alpha$ emission line of a spaxel clearly showing two velocity components 
(as evident from the double horn profile of the H$\alpha$ and [NII] lines) fitted with a single velocity component (top) and with two velocity 
components (bottom) by $QBfit$. 
The fit that accounts for just one velocity component finds a solution with a very large velocity dispersion ($\approx 125$ km s$^{-1}$).
In fact, a better fit is obtained when two velocity components are considered. We set the sigma threshold to discriminate when a double velocity fit may be needed to $35 \rm ~km~s^{-1}$, that is
the typical velocity dispersion of the gaseous component in galaxies at $z\sim0$ \citep{wisnioski15}. \\
After a double velocity component fit is performed,  $QBfit$ performs further checks in order to limit false or ambiguous detections:
the two components must be separated of at least $1.25 \AA$, which is the size of a pixel; both components must have S/N$> 5$ in order to represent a detection of a velocity component.
In each spaxel, if the solution does not satisfy these conditions or if the double velocity component fit does not converge, the result of the single velocity component fit is stored instead.
As a rule, $QBfit$ always stores the single velocity solution as part of the high velocity component.\\
To evaluate the stellar kinematics we perform a Gaussian fit to the second line of the CaII triplet ($\rm\lambda8542.09\AA$), after smoothing the datacube by 10x10 spaxels.
\section{Flux and Kinematics of gas and stars}
\label{sec_results}
\begin{figure*}
\centering
\includegraphics[width=17cm]{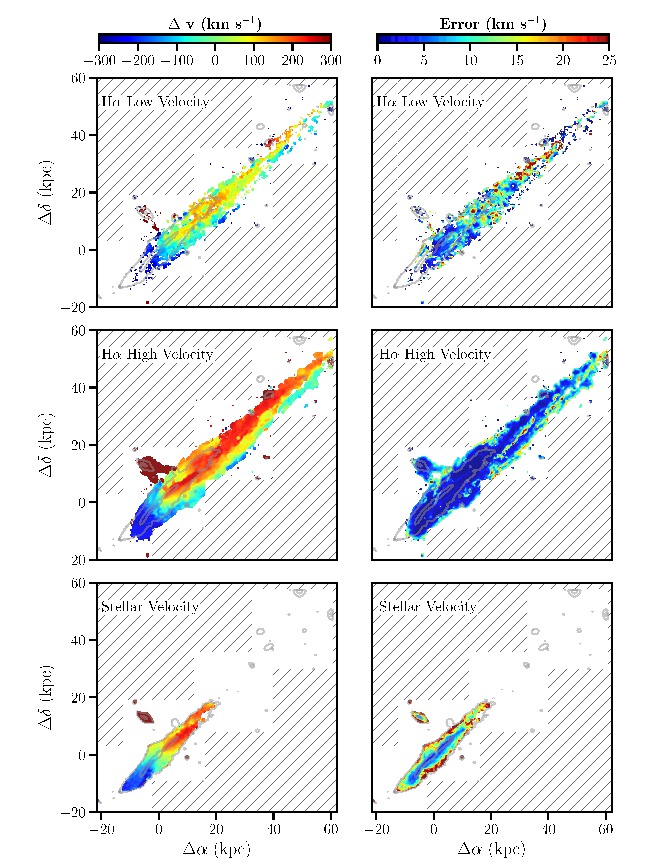}
\caption{The MUSE field  centered on UGC 6697 smoothed by 10x10 has been fitted with the H$\alpha$+[NII] lines assuming double velocity components. 
The first row shows the velocity field and the associated error map of the low velocity component of the gas.
In the second row, we present the velocity field and the associated error map of the high velocity component.
Finally, in the bottom row, we plot the stellar velocity map derived from the fit of the Ca II line and its associated error. 
The extent of the stellar continuum is given by the gray contours that represent the 23$^{th}$ 
mag arcsec$^{-2}$ and the 20$^{th}$ mag arcsec$^{-2}$ isophotes in the SDSS $r$-band image reconstructed from the MUSE datacube.
Notice that the SE tip of the galaxy does not show any line emission. 
}
\label{1compKin}  
\end{figure*}
\begin{figure}
\centering
\includegraphics[width=9cm]{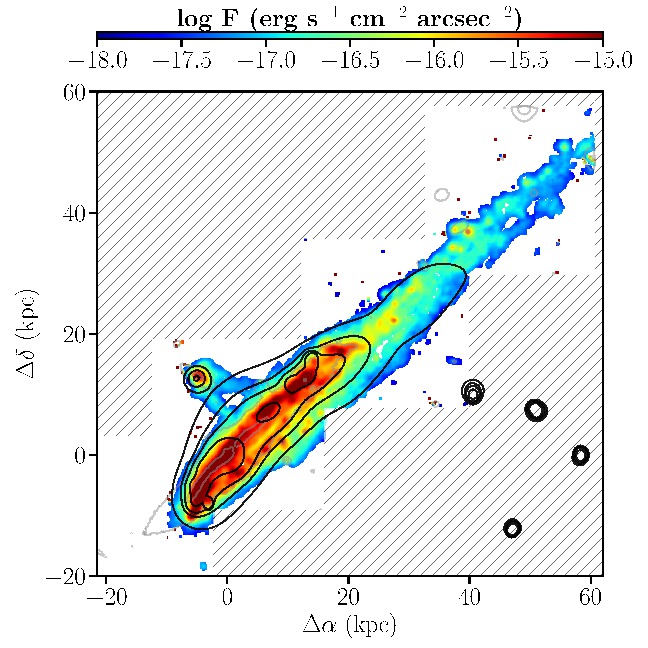}
\caption{Total flux distribution of H$\alpha$ emission with the Chandra contours superimposed to show the region of the X-ray emitting gas. The X-ray map has been smoothed, but the
X-ray front is at the same position of the H$\alpha$ front. The X-ray tail appears co-spatial to the H$\alpha$ tail.
}      
\label{xrayim}  
\end{figure}

In this section we discuss the kinematics of stars and gas as evaluated respectively from the Ca II absorption line ($\rm\lambda8542.09\AA$) and the H$\alpha$+[NII] emission lines,
exploring the results of the two velocity  component fit  
performed assuming the systemic redshift of UGC 6697 for the whole frame (hence CGCG 97807N is not centered at 0$\rm~ km~s^{-1}$).
In order to test the robustness of our double component fits, we compare to results obtained with KUBEVIZ \citep[see ][]{fox16}, finding  agreement.
The low-velocity and high-velocity maps, together with stellar kinematics, are shown in Fig. \ref{1compKin}.
The distribution of the spaxels that contain two velocity components is traced by the low velocity component, 
while the high velocity map contains the spaxel fitted with one velocity as well.\\
As it can be seen in the first two panels, the gas distribution of UGC 6697 extends in the NW direction for twice as much the extent of the stellar component 
traced by the grey contours of the continuum in the $r$ filter.
The extent and orientation of the tail suggests that UGC 6697 is traveling 
at high speed, almost in the plane of the sky, from NW to SE suffering the hydrodynamic interaction with the hot intacluster medium (ICM).
In the SE front of the stellar component of the galaxy, the gas is absent and no emission line is detected. The tail extends approximately for 50 kpc, 
including many HII regions (visible also in Fig. \ref{cube}), seen preferentially in the upper edge of the tail.
In Fig. \ref{xrayim} we show the X-ray contours of \citet{sun05} superimposed to our map of the total H$\alpha$ flux. The X-ray emission is coplanar to the H$\alpha$:
at the front, where the ISM impacts the ICM, the contours compress forming a sharp edge that corresponds to the one observed in H$\alpha$; 
in the NW back both tails overcome the length of the optical galaxy.
We will refer to the radius at which we observe such sharp edge as the observed stripping radius.
By considering the ICM density measured by \citet{sun05} ($\rm \rho_{ICM}\approx 4.5\times10^{-4} cm^{-3}$) at the position of UGC 6697 we can evaluate the ram pressure for different velocities.
On the other hand, we evaluate the anchoring force of UGC 6697  as a function of the radius as to be $\Pi = 2\pi G \Sigma_{gas}\Sigma_{stars}$ where $\Sigma_{gas}$ and $\Sigma_{stars}$ are modeled
with a simple exponential profile of the form $\Sigma = (M_d/(2\pi r_d^2))~e^{-r/r_d}$, where $M_d$ is the mass and r$_d$ the disk scale length.
We assume M$_{d,stars}=10.5$ $M_\odot$ and M$_{d,gas}=0.19\times M_{d,stars}$ (as evaluated from the HI mass measured by Gavazzi et al. 2001) and an r$_d \approx 5$ kpc, typical for a $10^{10.5}$ $M_\odot$ disk galaxy at z$\sim 0$ \citep{courteau07}.
By comparing the ram pressure obtained with different velocities to the anchoring force as a function of the radius, we can reproduce the observed stripping radius by considering a transit velocity for UGC 6697 of $\approx 500\rm~km~s^{-1}$.\\

The companion galaxy, CGCG 97087N, is losing gas while traveling almost perpendicular (from SW to NE) to the motion of UGC 6697 as suggested by the double tail 
emerging from the galaxy and connecting to the gas distribution of UGC 6697. 
Similarly to UGC 6697, also CGCG 97087N has a depleted front where we do not detect any gas emission but, differently from its massive companion, its tail splits in two and completely 
lacks HII regions. 
For UGC 6697, the integrated H$\alpha$ flux in the low velocity map is $1.16\times10^{-13} \rm erg~s^{-1}~cm^{-2}$ and $5.2\times10^{-13} \rm erg~s^{-1}~cm^{-2}$ in the high velocity frame.
CGCG 97087N instead  has an H$\alpha$ flux  as low as $6.9\times10^{-16} \rm erg~s^{-1}~cm^{-2}$ and $1.23\times10^{-14} \rm erg~s^{-1}~cm^{-2}$ in the low  and high velocity maps
respectively. 
The sum of the fluxes of the two component is in good agreement with the independent values published by Yagi et al. (2017), obtained from the narrow-band imaging with the 
Subaru telescope ($5.8\pm1.1\times \rm 10^{-13}erg~s^{-1}~cm^{-2}$).
\subsection{Velocities}
\label{sec_velocities}
\begin{figure}
\centering
\includegraphics[scale = 0.25]{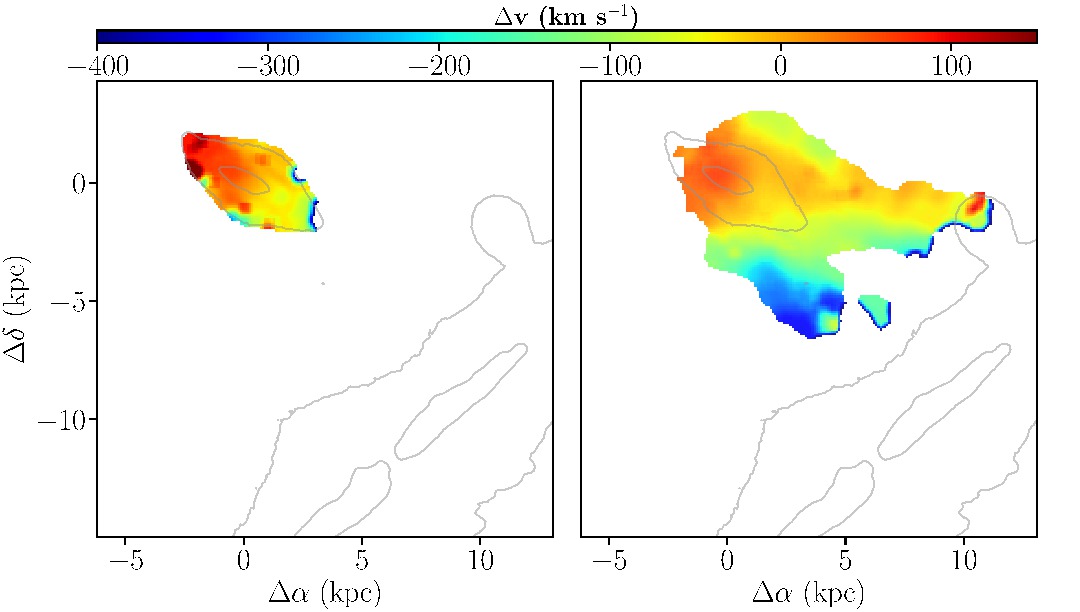}
\caption{Velocity maps centered on CGCG 97087N of the stellar component as derived from the Ca II lines (left) and of the gaseous component (right) as traced by the H$\alpha$. 
Gray contours represent the 23$^{th}$  and the 20$^{th}$  mag arcsec$^{-2}$  isophotes in the SDSS $r$-band image obtained from the datacube.
}
\label{87N_kin}  
\end{figure}

Fig. \ref{1compKin} shows the velocities of the ionized gas, separated in low and high velocity components, and of the stellar component along the line of sight after 
subtracting the systemic velocity of UGC 6697. The velocities of CGCG 97087N are displayed also in Fig. \ref{87N_kin}.
Both galaxies are seen nearly edge-on with clear evidence of a rotation curve in both the gas (first two rows) and the stars (bottom panel).
The rotation curve of stars as traced by the CaII triplet is smooth and fairly symmetric in both galaxies.
UGC 6697, the most massive of the two galaxies, rotates with a velocity of $\approx 250\rm~ km~s^{-1}$ (at a radius of $\approx 14$ kpc) while its smaller companion shows a rotational 
velocity of $\approx 100 ~\rm km~s^{-1}$ (at a radius of $\approx 3$ kpc). The rotation curve appears fairly regular and disfavouring a scenario in which the two galaxies 
have been interacting tidally. However, a mild
asymmetry is present in the NW edge of the stellar component of UGC 6697.
\begin{figure*}
\centering
\includegraphics[width=18cm]{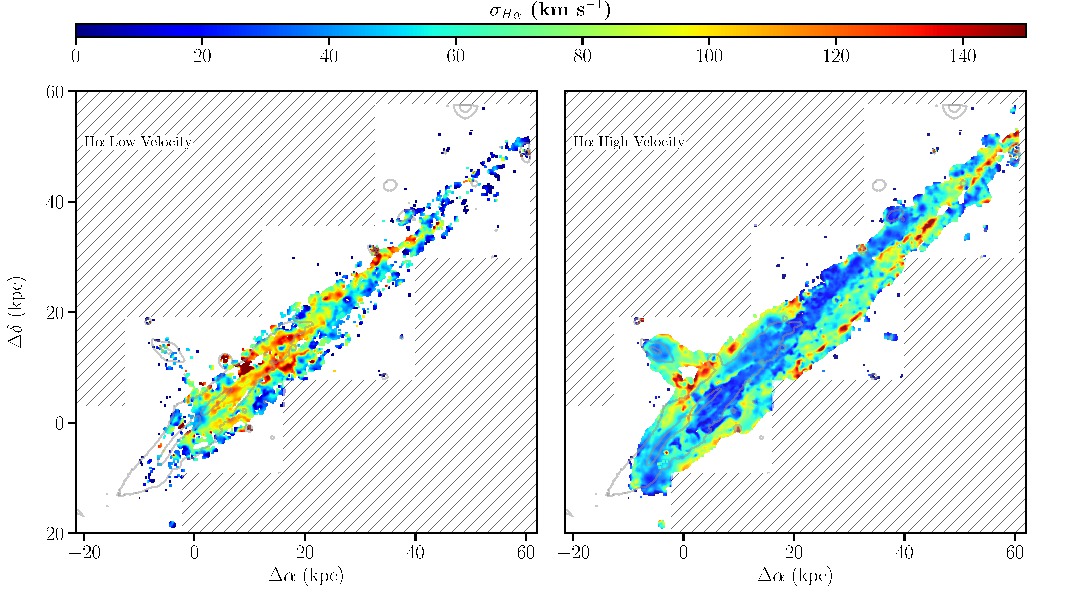}
\caption{Velocity dispersion map derived from the H$\alpha$ line, fitted with a double velocity component. In the left panel we plot the low velocity component
and in the right panel the high velocity component.
Gray contours represent the 23$^{th}$  and the 20$^{th}$  mag arcsec$^{-2}$  isophotes in the SDSS $r$-band image obtained from the datacube.
}
\label{gas_sigma1k}  
\end{figure*}
On the other hand, the interaction between the ISM and the ICM causes the gaseous component to leave UGC 6697 in the NW direction. 
Since the motion of the galaxy through the cluster is believed to be almost in the plane of the sky, the drag due to ram pressure does not decelerate the gas along the line of sight. 
In this geometry, the stripped gas would preserve its pre-stripping dynamical status. In other words the redshift 
of the stripped gas would remain close to the value it had before stripping, i.e. when it was attached to the stars, thus preserving memory of the stellar rotation curve. \\

In velocity space, the tail of UGC 6697 is divided in two substructures sharply separated by $\approx 200 \rm~km~s^{-1}$: the upper layer of the tail 
(in orange in the central map) connects to the NW edge of the stellar component
and its velocity is around $250\rm~km~s^{-1}$; the lower layer of the tail (in green/blue in the central map) is at much lower velocities (-100 - 0 $\rm~km~s^{-1}$) 
and connects to the central and SE-front part of the galaxy. Such a difference of velocities suggests that, similarly to CGCG 97087N, also the gas stripped 
from UGC 6697 is distributed along two tails that are seen almost in perfect superposition. In correspondence of the upper layer of the tail,
in the low velocity map, we  trace the underlying low velocity gas possibly associated to the lower layer of the tail.
All the HII regions that we detect in the NW tail are univocaly associated to the gas with the highest velocity that connects to the NW edge 
of the galaxy. 

The velocity dispersions of the two components are displayed in Fig. \ref{gas_sigma1k}.
Far in the NW tail, where the HII regions lies, the gas displays low velocity dispersions with respect to the surrounding gas, contrary to 
the low velocity gas which retains higher values of velocity dispersion suggestive of more turbulent kinematics.
The observed velocity dispersion of the gas outside the HII regions in the disk and in the tail is broadened up to
$\approx~150\rm km~s^{-1}$, i.e. approximately five times higher than the expected value for a spiral galaxy at $z\sim0$ \citep{wisnioski15}.
\begin{figure}[t]
\centering
\includegraphics[width=9cm]{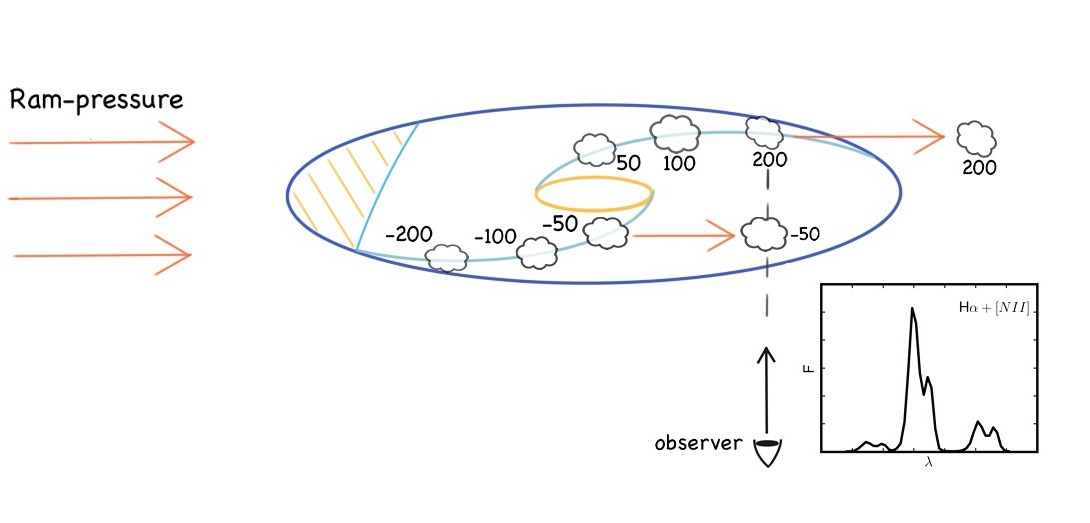}
\caption{Cartoon of UGC 6697 under the action of ram pressure due to its fast motion through the ICM. Its front (shaded) is already completely gas depleted.
The velocity of the gas is shown following the velocity curve of the stars. Velocities are given normalized to zero in the center.
Some gas originally belonging to a central region has been displaced to the NW by ram pressure, retaining its original velocity (-50 km/s).
This cloud overlaps spatially with another gas cloud on-board the background arm, with a velocity of 200 km/s. An observer who looks 
at the galaxy almost edge-on
collects a spectrum carrying multiple velocity components along the NW tail.  
}
\label{cartoon}  
\end{figure}

Figure \ref{cartoon} sketches our cartoon model for UGC 6697, including the presence
of double velocity components. In this model, the galaxy is in fast motion through the ICM and is observed along an almost edge-on view.
Its front (shaded) has been already completely depleted of gas. Indeed the presence of
PSB spectral features in this region testifies an abrupt truncation of the star formation. 
As the line of sight approaches the central region, we detect gas at increasing velocity 
following the rising rotation curve of the stars. 
More to the NW, most lines of sight will cross regions containing multiple velocities: those "on board" 
the stellar rotation curve (marked 200) are blended with those containing stripped gas (marked -50) that originate
in the SE and retain the kinematics of their places of origin. 
The spectrum (inset of Fig. \ref{cartoon}) shows a typical
double horned profile of H$\alpha$ and [NII] with separation exceeding 200 $\rm km~s^{-1}$, as observed mainly in the down-stream half of UGC 6697.

Fig. \ref{87N_kin} shows the velocity field of CGCG 97087N for its stellar and gaseous component separately. For the gaseous component, we plot the
high velocity component since we detect only few spaxels in the low velocity component, representing a negligible percentage of the flux belonging to CGCG 97087N.
The map shows the gas being dragged out of the galaxy by the ram pressure from NE to SW.
Its lower rotational speed is consistent with the low mass measured by Consolandi et al. (2016) and the high recessional velocity with respect to the cluster suggests 
a motion through the cluster with a significant component along the line-of-sight.\\

The stellar component rotation curve is fairly smooth with only a mild velocity gradient. 
On the contrary the map of the gas looks rather different.
The ionized gas is stripped along two tails  that are separated spatially and kinematically: the southern (S) tail and the northern (N) tail. 
Along the N tail we detect velocities that are very similar to the velocities of the stars (between -100 and 100 $\rm~km~s^{-1}$). 
\begin{figure}
\centering
\includegraphics[width=9cm]{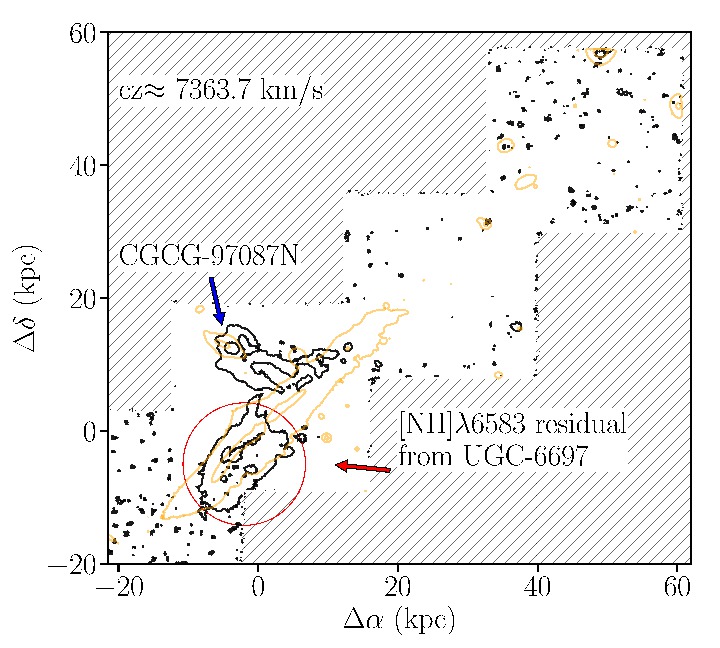}
\caption{Contour of the H$\alpha$ flux of the tails of UGC 6697 and CGCG 97087N once that the [NII] emission has been isolated and subtracted. The red circle indicates the
contours arising from the residual flux of the [NII]$\lambda6583$ emitted by UGC 6697. Regions not observed are shaded in gray. Yellow contours represent the 23$^{th}$  and the 
 20$^{th}$  mag arcsec$^{-2}$  isophotes in the SDSS $r$-band image obtained from the datacube.}      
\label{codine}  
\end{figure}
Oppositely, the S tail shows a much steeper velocity gradient: the velocity of the gas from the S edge of the galaxy continuously 
decreases down to $\approx$ -400 $\rm~km~s^{-1}$. Such gradient is not consistent with the velocities displayed by the velocity curve of the stars. 
Instead, such a significant deceleration of the gas could be caused by the drag induced on the ISM by the cluster ICM. 
In this case, however, a similar deceleration should appear in the N tail too.
However we tend to exclude that the gas that we observe at -400 $\rm~km~s^{-1}$ belongs instead to UGC 6697, which would support a possible direct interaction between the two galaxies. 
There is in fact a gap of more than 500$\rm~km~s^{-1}$ between the
lowest velocity in the N tail and the velocity of the gas unambiguously belonging to UGC 6697 observed near the region of putative crossing. 
Nevertheless, the flux of both tails eventually connect to the gas distribution of UGC 6697 (see also Fig. \ref{1compKin}) but there is no evident continuity 
in the velocity distributions.
However, looking at Fig. \ref{gas_sigma1k} we note that the dispersion in the tails of CGCG 97087N seem to form a continuum with the high velocity dispersion gas belonging to UGC 6697 in its upper periphery, 
suggesting a physical connection between the two galaxies.\\
It is therefore interesting to test the possibility that a fast hydrodynamic interaction occurred  between the two galaxies, possibly 
enhancing the ram pressure acting on CGCG 97087N. 
Signal from the tails of CGCG 97087N is detected from 7245 to 7647 $\rm~km~s^{-1}$.
In absence of a precise 3-D distance indicator, it is impossible to establish if
the gas trailing behind CGCG 97087N is just projected on UGC 6697 (either on the foreground or on the background) or has indeed crossed the disk of the main galaxy.
If projected, it would indicate that CGCG 97087N is under the action of ram pressure stripping by the ICM of Abell 1367. Conversely, gas would be stripped by the combined
action of the ICM and the ISM of UGC 6697. In either cases, CGCG 97087N  would be a member of Abell 1367 and not a background object.\\

In order to find evidence of a past interaction between the two galaxies we need to detect gas that is deeply embedded in the signal from UGC 6697.
In principle, once that the continuum is properly subtracted, the difference in redshifts should 
automatically separate the fluxes associated with the two galaxies. Unfortunately, the difference in redshift is such that the H$\alpha$ emission
of the tails of CGCG 97087N falls at the wavelengths of the [NII]$\lambda6583$ line of UGC 6697. Moreover the tails cross each other at the position
where the emission from UGC 6697 is at its maximum, so that the [NII]$\lambda6583$ line outshines the faint signal of the H$\alpha$ line belonging to the tail of CGCG 97087N.
In these regions the signal of UGC 6697 is also characterized by a double velocity component that is brighter than the signal from CGCG 97087N, hampering 
the detection of a putative third component.\\
In order to overcome this problem, we use a non-parametric method to model and subtract the continuum and isolate only the emission of the H$\alpha$ line. 
In each individual spaxel, the procedure starts from the first element on the wavelength dimension and, for each spectral step, it derives a redshift ($z$) assuming that the flux (F($\lambda$)) 
at each wavelength ($\lambda$) is [NII]$\lambda6548$ that has a flux of $1/3$ of the flux of the second line of the [NII] at $\lambda6583$.
Under these assumptions, the code examines the flux at $\lambda_{2}=\lambda+\Delta\lambda_{[NII]}\times(1+z)$ and if F($\lambda_{2})>3\times$ F$(\lambda)$, the procedure flags
emission as [NII]$\lambda6583$. The procedure then subtracts the F$(\lambda)$ and, at wavelength $\lambda_{2}$ subtracts $3\times$F$(\lambda)$.\\ 
This procedure successfully recognizes the [NII] lines and the subtraction is satisfactory, although not perfect especially in the wings of the emission line. 
In the regions where the [NII] lines are very bright the subtraction is performed at $5-10\%$ percent level.
In Fig. \ref{codine} we show the contour map of the channel at the  velocity where the H$\alpha$ of CGCG 97087N is found.
Surface brightness contours are shown in steps of $\times10$ starting from $\approx 3.3\times10^{-19}\rm erg~s^{-1}~cm^{-2}~arcsec^{-2}$. 
Intriguingly, the subtraction of the [NII]  lines leaves a faint signal consistent with H$\alpha$ emission belonging to CGCG 97087N elongating along the NW direction on the main body of UGC 6697.
The appearance of this faint but well defined double tail naturally connecting with the emission observed outside the bright signal of UGC 6697 suggests
that this embedded signal is real, and that CGCG 97087N interacted hydrodynamically with UGC 6697.\\

In order to quantify the enhancement of the ram pressure acting on CGCG 97087N due to the ISM of UGC 6697, we evaluated the anchoring force of CGCG 97087N as a function of the radius ($\Pi = 2\pi G \Sigma_{gas}\Sigma_{stars}$ ) and
compared it to the possible value of ram pressure due to the interaction with diffferent phases of the ISM assuming that the difference in the systemic velocities of the two galaxies 
is the relative velocity between the galaxy and the ISM. Similarly to what has been done for UGC 6697 we modeled the stellar and gaseous disk with 
the same exponential function considerng M$_{d,stars}=10^{9.1}$ $M_\odot$, a gas fraction of 1 and a disk scale length of 1 kpc.

We assume the same ICM density assumed for UGC 6697 and a recessional velocity of 990 $\rm~km~s^{-1}$, the ram pressure felt by CGCG 97087N is P$_{ram} = 7.4\times10^{-12} \rm ~dyn~cm^{-2}$
We then consider the three phases of the ISM with the greatest filling factor: the hot ionized medium \citep[n$\approx \rm 10^{-3} ~cm^{-3}$][]{ism}, which is thought to pervade much of the volume above and below the disk;  
the warm ionized medium  (n$\approx \rm 0.1 ~cm^{-3}$) and the warm neutral medium \citep[n$\approx \rm 1 ~cm^{-3}$][]{ism}.
Considering a relative velocity of 870 $\rm~km~s^{-1}$ between CGCG 97087N and UGC 6697 the ram pressure due to the dense warm ionized and neutral medium overcomes the anchorig force of CGCG 97087N by orders of magnitude at all radii, possibly depleting it almost instantaneously.
However, at the densities of the hot coronal ionized medium, the ram pressure would be only mildly enhanced during the short crossing time ($\approx 0.1 - 1$ Myr considering a gas disk scale height between 0.1 and 1 kpc), enough for the galaxy to retain part of the gas and at a stripping radius
consistent with the observed one. Hence CGCG 97087N possibly did not interacted with the densest phase of the ISM while crossing the disk but possibly interacted with the less dense hot halo of UGC 6697.

\subsection{Onboard and stripped gas}
\label{sec_2vel}
\begin{figure*}
\centering
\includegraphics[width=18cm]{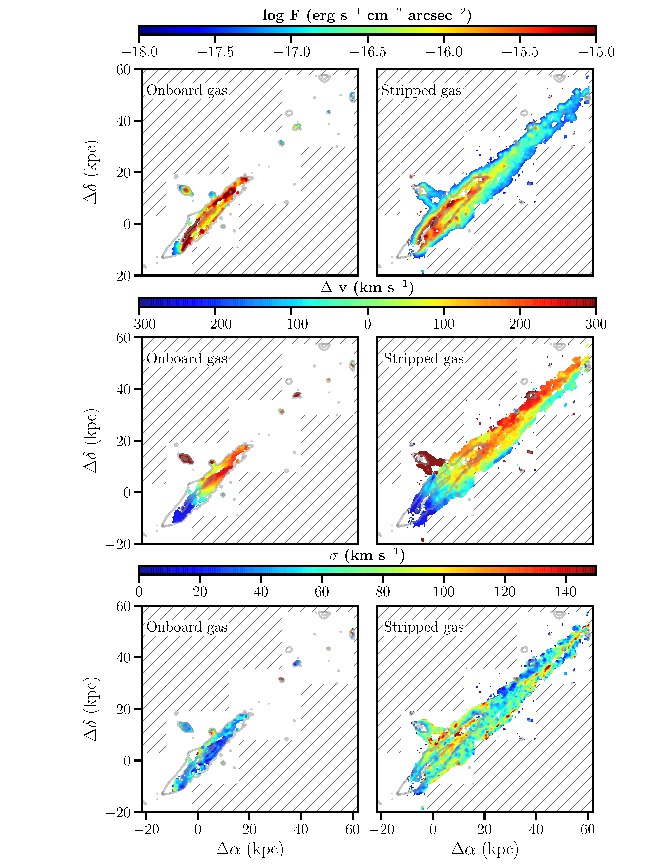}
\caption{The MUSE field centered on UGC 6697 smoothed by 10x10 pixels has been fitted with two sets of the H$\alpha$+[NII] lines with different velocities along the line-of-sight. 
After adopting a $\delta v$ threshold of 75 $~\rm km~s^{-1}$ between gas and stars to discriminate the gas emission still bound to the galaxy and the stripped gas emission,
we separate the onboard gas (left column) and the stripped gas (right column).
The flux, velocity and velocity dispersion maps are displayed from top to bottom and 
gray contours and the cross in each panel are the same as in previous figures.
}
\label{stripon}  
\end{figure*}
By comparing the gas properties of both the low and the high velocity component with the kinematics of the stars we attempt to separate the flux and
kinematics of the stripped gas (dominated by the action of ram pressure) from the one that is possibly still onboard the galaxy. First we assume that the gas in projection outside the stellar distribution (as traced by the CaII lines) 
is stripped by definition. Second, we adopt a threshold of $\Delta v = |v_{gas} - v_{stars}|= 75~\rm km~s^{-1}$ 
in each spaxel for the gas emission to be considered either bound to the galaxy ($\Delta v <75 ~\rm km~s^{-1}$) or stripped ($\Delta v >75 ~\rm km~s^{-1}$)
\footnote{In spaxels where both the velocity components have a $\Delta v$ greater than $75~\rm km~s^{-1}$ with respect to the stellar component we consider both belonging to stripped gas and we
take into account the sum of the fluxes, and we average the velocities and the velocity dispersions}.
Such threshold has been chosen to represent the quadratic sum of the maximum stellar and gas velocity dispersion expected 
in local disk galaxies \citep[$\approx 50~\rm km~s^{-1}$][]{kregel05,wisnioski15}. It is worth stressing that our criterion based on the gas kinematics and 
geometry does not separate gas that is gravitationally bound to the gas that will be lost in the ICM. Instead, some of the gas close to the galaxy that we considered stripped 
may eventually fall back once the ram pressure becomes negligible.

Fig. \ref{stripon} show the result of our attempt to separate the stripped gas from that remaining in the galaxy.
We estimate that the flux still attached to UGC 6697 is 
$\approx3.2\times10^{-13} \rm erg~s^{-1}~cm^{-2}$, comparable with the stripped
gas that has a flux of $\approx2.8\times10^{-13} \rm erg~s^{-1}~cm^{-2}$. As for CGCG 97087N, the flux splits in
$\approx9.2\times10^{-15} \rm erg~s^{-1}~cm^{-2}$  for the gas still associated with the galaxy and $\approx3.63\times10^{-15} \rm erg~s^{-1}~cm^{-15}$ that is stripped away.
The inboard gas is associated with the brightest regions of the continuum (see contours) and with the NW side of the galaxy.\\
The second and third rows of Fig. \ref{stripon} show the kinematics (velocity and dispersion) as traced by the H$\alpha$ line. 
It is remarkable that such a simple separation based solely on the velocity difference between stars and gas, with no further assumptions
on the velocity dispersion, leads naturally to separate a component dominated by low velocity dispersions from a component characterized by a higher average velocity
dispersion. 
Indeed, the gas considered onboard has a dispersion  
comparable to the expected values of \citet{wisnioski15}, while the gas considered to be stripped shows much higher velocity dispersion.
The only exception is in the upper layer of the tail, where most HII regions in the tail are found. This region is characterized instead by low velocity dispersion
(blue) similar to the gas bound to UGC 6697.
Once again the velocity dispersion map of the tails of CGCG 97087N (bottom, right in Fig. \ref{stripon}) displays a structure that connects to the gas 
of UGC 6697, again suggesting a physical connection between the two galaxies.\\

Once the stripped gas has been separated from the gas that remains bound to the galaxy, we quantify the mass of ionized gas 
that has been ripped from the galaxy. This can be estimated from the electron density by assuming a geometry for the  H$\alpha$ emission of the stripped gas.
The electron density can be derived from the H$\alpha$ luminosity using the relation 
\begin{equation}
L(H\alpha) = n_e n_p\alpha^{eff}_{H\alpha}Vfh\nu_{H\alpha}
\end{equation}
\citep{oster06}, where $\rm n_e$ and $\rm n_p$ are the number density of electrons and protons, $\alpha^{eff}_{H\alpha}$ is the H$\alpha$ effective recombination 
coefficient, V is the volume of the emitting region, $f$ is the filling factor, $h$ is the Planck's constant, and $\rm \nu_{H\alpha}$ is the frequency of the H$\alpha$ emission line.
Unfortunately this estimate is affected by the high degree of uncertainty in some parameters. In particular,
the exact geometry of the tail is unknown and we can only approximate its structure assuming that the emitting region has a cylindrical shape of 
diameter equal to the height of the galaxy and length of $\approx$90 kpc. Because
the morphology of the tail  hints at a filamentary distribution containing high density clumps and compact knots, a filling factor lower than 1 is assumed.
In particular, we assume $f=0.1$, consistently with previous works that generally consider $0.05<f<0.1$ \citep{fox16,1690}. Moreover, we assume that the gas is fully ionized,  thus implying
$n_e=n_p$ and  $\alpha^{eff}_{H\alpha} = 1.17\times10^{-13}\rm~cm^3s^{-1}$ \citep{oster06}.
Assuming a distance of 94.8 Mpc, the resulting electron number density is of the order of 0.6 cm$^{-3}$ which is in agreement with the density estimate by \citet{yagi07}, and \citet{fox12,fox16}.
The implied mass of ionized gas dragged outside the galaxy is of $\approx 10^{9.1}\rm M_\odot$ and we estimate its recombination time to be $\approx0.2$ Myr using  
\begin{equation}
\tau_r=\frac{1}{n_e\alpha_A},
\end{equation}
where $\rm\alpha_A$ is the total recombination coefficient equal to $\rm4.2\times10^{-13}~cm^3s^{-1}$ \citep{oster06}. 
The estimated stripped mass is consistent with the gas mass retained at radii $r\gtrsim 10$ kpc (the observed stripping radius) in a $10^{10.5}$ $\rm M_\odot$ disk galaxy modeled with an exponential function in a similar way as in the first paragraph of section \ref{sec_results}.  \\

By considering the projected separation between the stripped gas at lowest velocity in the NW tail and the region of the galaxy where stars have similar velocity,
we estimate a timescale for gas stripping. Assuming that the gas is instantaneously accelerated to the cluster velocity,
we estimate that these plumes have been stripped at least {\bf $\approx 60-120 $ Myr} ago, assuming that the galaxy travels at {\bf $\approx 500-1000~\rm km~s^{-1}$} in accordance
with the upper limit for its velocity estimated by \cite{sun05}.
Comparing this stripping timescale to the recombination time we conclude that some excitation mechanisms is necessary to keep the gas of the tail ionized.\\
Moreover, in the model proposed by \citet{Nulsen82} and \citet{sun05}, given the orientation and geometry of the impact with the ICM of UGC 6697, 
mechanism such as viscous stripping and Kelvin-Helmholtz instabilities are thought to dominate the stripping of the gaseous component.
Considering the density of the ICM $\rm \rho_{ICM}=4.5\times10^{-4}~cm^{-3}$ \citep{sun05}, we can evaluate the time necessary
to strip $\approx 10^{9.1}\rm M_\odot$ of ionized gas by Kelvin-Helmholtz instabilities if the galaxy travels through the ICM at {\bf $\approx 500-1000~\rm km~s^{-1}$} using:
\begin{equation}
\dot{M}_{K-H} \approx \pi r^2 \rho_{ICM} v_{gal},
\end{equation} 
\citep{Nulsen82}.
The mass of ionized gas in the tail can therefore easily be stripped within $\rm\approx7-14\times 10^7~yr$, which is consistent with the previous estimate of $\approx60-120$ Myr.
\section{Excitation mechanisms}
\begin{figure*}
\centering
\includegraphics[width=18cm]{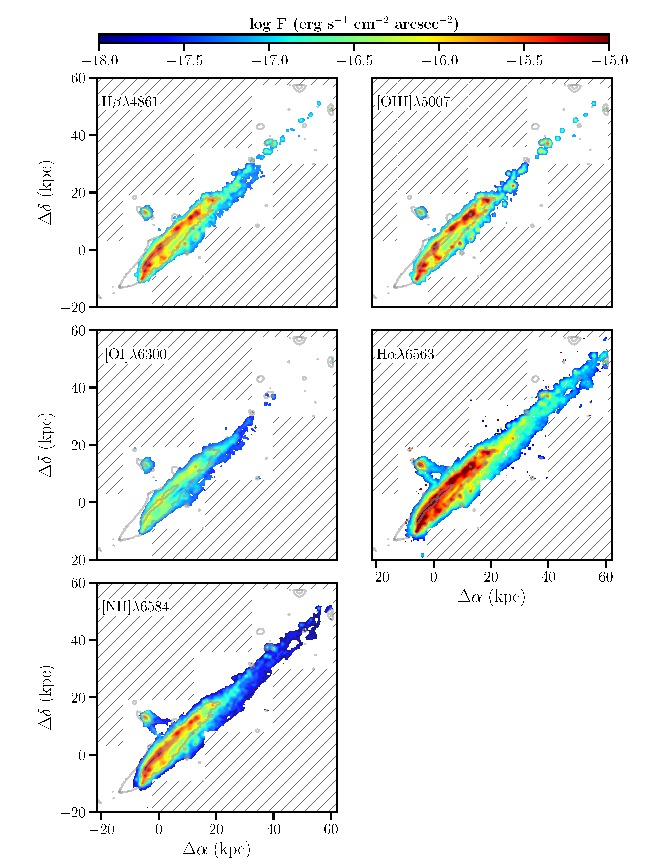}
\caption{Emission line maps of UGC 6697. Panels are sorted by increasing wavelength from left to right and from top to bottom. The solid contours 
represents the 23$^{th}$ and the 20$^{th}$ mag arcsec$^{-2}$ isophotes in the SDSS $r$-band image obtained from the datacube.
Areas not mapped by the MUSE mosaic are shaded in grey.  
}      
\label{allflux}  
\end{figure*}
To constrain the ionizing processes that are acting on the gas of the tail, we consider ratios of emission lines.
In particular, we focus our attention on the  BPT diagnostic diagram \citep{bpt}, considering lines that fall within the MUSE spectral 
window: H$\beta\lambda4861$, [OIII]$\lambda5007$, [OI]$\lambda6300$ and [NII]$\lambda6584$.
The double velocity component fit to each lines keeps fixed the velocity and velocity dispersion of the two velocity components to the values evaluated 
during the fit of the H$\alpha$ line. This ensures that all lines have been fitted with consistent kinematics fixed by the
best resolved and brightest line available.
The maps showing the distribution of the fluxes of all emission lines that we fitted are displayed in Fig. \ref{allflux}.
These fluxes are the sum of the two velocity components.
The [SII] lines are not considered because they fall at the wavelength of a strong sky emission line, preventing a robust flux determination.\\
Emission lines are detected along the main body of the galaxy while in the outer tail 
the distributions differ significantly.
Hydrogen (H$\alpha$ and H$\beta$) is found in both the diffuse gas and HII regions along the whole extent of the tail.
The [OIII]$\lambda5007$ line is preferentially found in the HII regions of the tail and only little flux belongs
to diffuse gas emission; conversely [OI]$\lambda6300$ is fainter and appears less important in HII regions.

Furthermore, following the same criteria that we adopted to separate the onboard and stripped H$\alpha$, we split each line in an onboard and a stripped component.
\subsection{Line ratio maps}
\label{sec_chemis}
\begin{figure*}
\centering
\includegraphics[width=18cm]{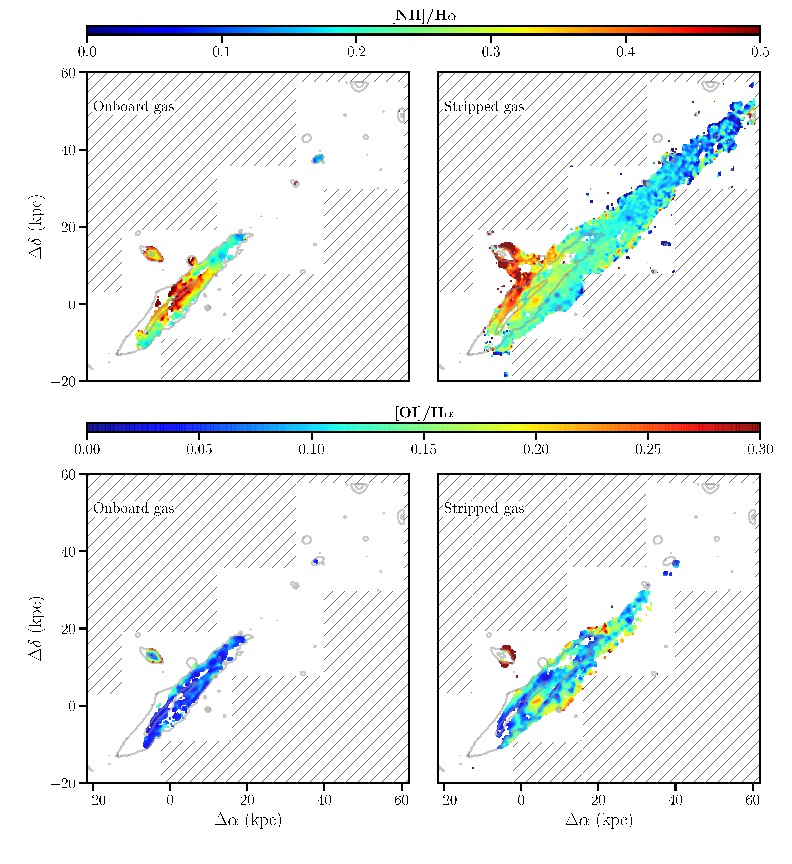}
\caption{Maps of line ratios for the onboard (left) and stripped (right) components. From top to bottom, we plot the [NII]$\lambda6584$/H$\alpha$ and [OI]$\lambda6300$/H$\alpha$ maps, 
respectively. Contours and gray shaded areas are the same as in previous figures. 
}      
\label{ratios2comp}  
\end{figure*}

\begin{figure*}
\centering
\includegraphics[width=18cm]{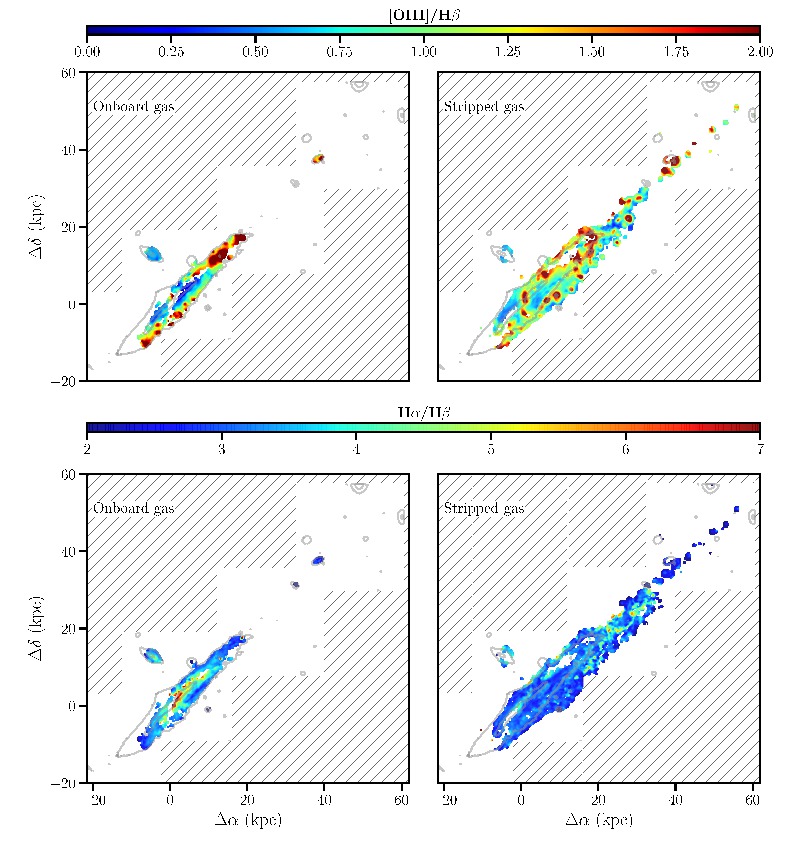}
\caption{Same as Fig. \ref{ratios2comp} but for the
[OIII]$\lambda5007$/H$\beta$ and H$\alpha$/H$\beta$ maps, respectively.}      
\label{ratios2comp2}  
\end{figure*}
Line ratio maps are displayed in Fig. \ref{ratios2comp} and Fig. \ref{ratios2comp2} for the onboard and stripped gas, respectively.
The top-row of Fig. \ref{ratios2comp} shows the [NII]$\lambda6584$/H$\alpha$ maps. This ratio maintains modest values ($\sim 0.2-0.3$) on the main body of
CGCG 97087N, while in the central region of UGC 6697 the ratio is slightly enhanced with respect to the rest of the galaxy. 
The value varies significantly also in the CGCG 97087N tails where it is greatly enhanced and in the tail of UGC 6697 where it drops to 
values lower than 0.2. This ratio is enhanced by shocks \citep{kew01,allen08} and, in our map of the onboard gas, 
the highest values are found along an elongated central region of the main body of UGC 6697 and at the surface of contact where the tails of CGCG 97087N point.
In the stripped gas maps, the highest values are in the wings and in the tails connecting CGCG 97087N to the galaxy.
The gas characterized by high values of the  [NII]$\lambda6584$/H$\alpha$ ratio connects and mixes to the gas belonging to UGC 6697, suggesting
once again a physical connection between the gas of the two galaxies.
On the contrary, in the tail of UGC 6697 the ratio drops to values below 0.2.

A major difference between the gas that we consider onboard the galaxy and 
the gas stripped in the tails is visible in the [OI]$\lambda6300$/H$\alpha$ map (second row of Fig. \ref{ratios2comp}), another
indicator of the presence of shocked gas. The gas belonging to the main body of both galaxies and to HII regions show very low ratio.
On the contrary, the gas outside the two galaxies is characterized by much higher values, reaching the maximum in the two wings of CGCG 97087N and in the surface connecting to its tails.
Similarly to what  has been observed in the [NII]$\lambda6584$/H$\alpha$ map, this strongly hints at  hydrodynamic interactions between the two systems. \\

The [OIII]$\lambda5007$/H$\beta$ map is in the first row of Fig \ref{ratios2comp2}.
The highest values of the ratio in both maps are found preferentially in regions associated to the bright HII regions visible both in the H$\alpha$
distribution and in the $r$-band image in the south direction and in the NW tail. 
This is consistent with an higher ionization factor (assuming a constant metallicity in the stripped gas) in those regions with respect to the surrounding gas.
The main body of both UGC 6697 and CGCG 97087N show onboard components with moderate [OIII]$\lambda5007$/H$\beta$ values with the exception of the peripherical layer
of UGC 6697, where HII regions lie.
This is also true for the SE edge, in correspondence with the sharp cut in the ionized gas distribution 
possibly associated to the front of the galaxy where the ISM impacts the hot ICM. In this case, we are probably looking at an enhanced star formation activity 
due to the compression of the gas in the front.\\

Finally we show the H$\alpha$/H$\beta$ map in the bottom row of  Fig. \ref{ratios2comp2}.
This ratio is commonly used to infer dust extinction, assuming an intrinsic ratio of 2.86 from case B recombination at $\rm T=10^4 K$ \citep{oster89}.
In the onboard map, the main body of UGC 6697 displays overall ratios that are consistent with being absorbed by the presence of dust.
On the contrary the stripped gas is in line with the theoretical value of the ratio. 
Also the main body of CGCG 97087N is characterized by higher ratio values ($\approx 4$) consistent with a mild obscuration by dust.
We stress that also UGC 6697 has higher values of H$\alpha$/H$\beta$ where the continuum is at a maximum, while in the tail the ratio settles around the theoretical value.
The presence of dust along the line of sight is not surprising  in the disk of edge on spiral galaxies but
we refrain from interpreting the values in the tails, because other ionization mechanisms are at play 
and the assumption of an intrinsic ratio of 2.86 may not hold true in these regions.
\subsection{Spatially resolved BPT diagrams}
\label{sec_bpt}
\begin{figure*} 
\centering
\includegraphics[width=18cm]{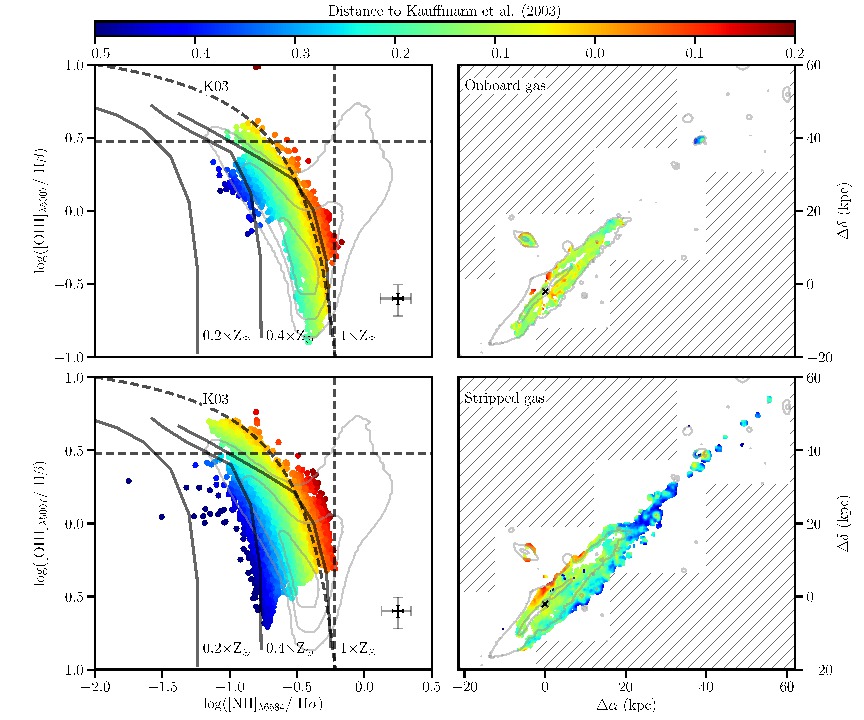}
\caption{{\bf Top.} Left: BPT diagram of the onboard component of UGC 6697. The dashed curve separates 
AGN from HII regions and is from \citet[][ K03]{kauff03}. Data  are color coded according to their minimum distance
to the K03 curve. The black and gray crosses indicate the typical error of the ratio of lines with S/N$\approx15$ and S/N$\approx5$, respectively.
At higher S/N ratios, the error becomes comparable to the dot size. 
Thick solid lines show three different photo-ionization models at different metallicities ($0.2$, $0.4$, $1$ Z$_{\odot}$) by \citet{kew01}. 
The grey contours are obtained from a random sample of nuclear spectra of SDSS galaxies in the redshift range 0.01 - 0.1 and with masses
from 10$^{9}$ and 10$^{11}$ M$_{\odot}$. 
Right: Map  of UGC 6697 of the spaxel contributing to the BPT of the onboard component, color coded as in the left panel. The black cross indicates the center of the galaxy while 
regions of the sky not mapped by the MUSE mosaic are shaded in gray.
{\bf Bottom.} Same as in the top panels but for the stripped component of the gas.}      
\label{bpt}  
\end{figure*}
\begin{figure*} 
\centering
\includegraphics[width=18cm]{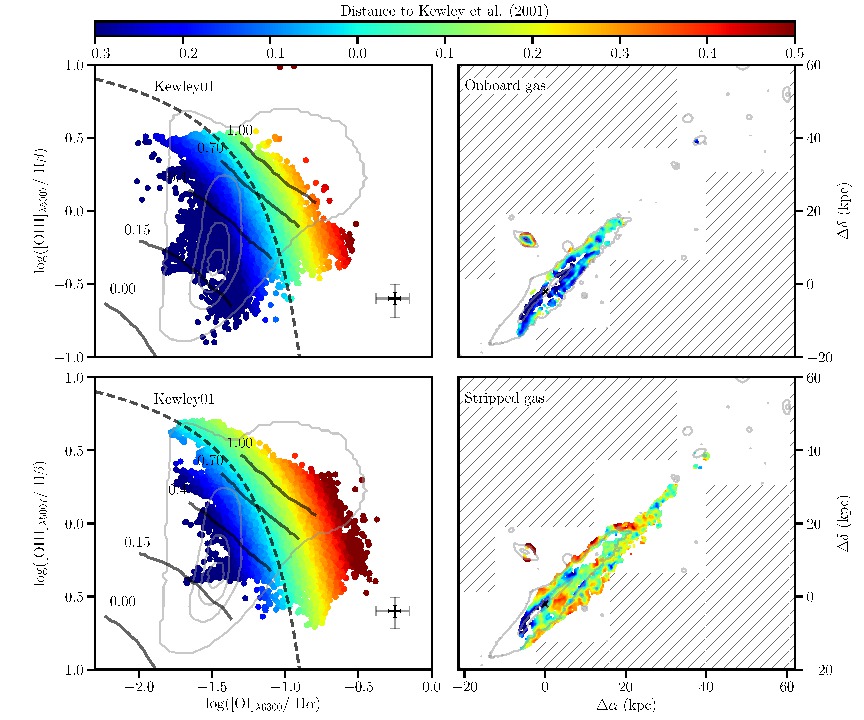}
\caption{{\bf Top.} Left: BPT diagram of the onboard component of UGC 6697 using the [OI] to H $\alpha$ ratio. The dashed curve separates 
AGN from HII regions and is  from \citet[][kewley01]{kew01}. Data are color coded according to their minimum distance
to the Kewley01 curve. The black and gray crosses indicate the typical error of the ratio of lines with S/N$\approx15$ and S/N$\approx5$, respectively.
At higher S/N ratios, the error becomes comparable to the dot size. Thick solid lines show five different shock models by \citet{rich11} indicating five different fractions (from 0 to 1) 
of H$\alpha$ flux contributed by shocks. 
The grey contours are obtained from a random sample of nuclear spectra of SDSS galaxies in the redshift range $0.01$-$0.1$ and with masses
from 10$^{9}$ and 10$^{11}$ M$_{\odot}$.
Right: Map of UGC 6697 of the spaxel contributing to the BPT of the onboard component, color coded as in the left panel. The black cross indicates the center of the galaxy while 
regions of the sky not mapped by the MUSE mosaic are shaded in gray.
{\bf Bottom.} Same as in the top panels but for the stripped component of the gas.}      
\label{bptoi}  
\end{figure*}
Next we build spatially resolved BPT diagrams for the onboard and stripped gas, separately.
Unfortunately, the required simultaneous detection of four emission lines in each spaxel limits the analysis to the brightest spaxels and we cannot use the BPT diagnostic along
the whole extent of the tail.\\
In the left panels of Fig. \ref{bpt} we show
the BPT diagrams based on the H$\beta\lambda4861$, [OIII]$\lambda5007$, H$\alpha\lambda6563$ and [NII]$\lambda6684$ emission lines of the onboard and stripped component, respectively.
Resolution elements are color coded according to their minimum distance from the \citet{kauff03} curve that separates the regions of the plane associated to photo-ionization 
from the regions associated to other ionization processes (e.g. AGN and shocks). 
For each BPT panel, on the right, we display the map color-coded in the same way as the associated BPT digram in order to locate which region of the galaxy contribute to
which points in the BPT digram.
In both panels, we overplot three different photo-ionization models at different metallicities ($0.2$, $0.4$, $1$ Z$_{\odot}$) from \citet{kew01}. The contours are
obtained from a sample of nuclear spectra of SDSS galaxies in the redshift range $0.01<z<0.1$ and with stellar masses $\rm M_*=10^{9}-10^{11} M_{\odot}$.
In the onboard component, our data mainly distribute along the HII region wing of the BPT: both the main body of UGC 6697 and the main body of CGCG 97087N are characterized 
by typical ratios for gas photoionized by stars \footnote{The few blue points in the BPT of the onboard component are not from UGC 6697 but, instead, from the small galaxy at x$\approx40$ kpc y$\approx40$ kpc. }.
The spaxels associated to the diffuse gas outside the galaxy (bottom panels of Fig. \ref{bpt}) exhibit instead a lower [OIII]$\lambda5007$/H$\beta$ ratio compared to HII regions inside the galaxy. 
This deviation from the HII regions wing of the BPT can be interpreted as an effect of the lower metallicity of the peripherical gas of the galaxy. The external gas is indeed less bound to the galaxy and
stripped more easily. The photo-ionization model with Z$\rm=0.4Z_{\odot}$ of \citet{kew01} falls perfectly within our data.
The (red) points in the BPT that lies just at the right of the curve of \citet{kauff03} are associated with  the wings of CGCG 97087N and with the upper layer of 
UGC 6697, where the tails of CGCG 97087N intersect UGC 6697.\\

In Fig. \ref{bptoi}, from top to bottom, we plot the BPT diagnostics based on H$\beta\lambda4861$, [OIII]$\lambda5007$, H$\alpha\lambda6563$ and [OI]$\lambda6300$ for 
the onboard and stripped component of the gas, respectively.  Points and maps are color coded according to the minimum distance from the \citet{kew01} curve that limits the
region consistent with photoionization. 
We overplot  five curves indicating the relative contribution of shocks to the ionization of the gas \citep{rich11}. 
The contours are again obtained from the same sample of nuclear spectra of SDSS galaxies previously described.
In the top panel, the onboard gas is distributed mainly in the photoionization region of the diagram and the only points crossing the limit of \citet{kew01} (dashed curve) 
are associated to the outer parts of CGCG 97087N and to peripherical regions of UGC 6697.  
A clear difference is observed instead in the diagram of the bottom panel: the majority of the points fall in the region at the right of the limit of 
photoionization of \citet{kew01}. \\

According to the models of \citet{rich11}, roughly half of spaxels are consistent with $70\%$ of the ionization coming from shocks. 
The regions with the highest contribution from shocks (up to 100$\%$) are associated with the diffuse gas at the periphery of the galaxy, while the
main body of both galaxies appear consistently photo-ionized by stars.\\

Altogether, Figs. \ref{bpt} and \ref{bptoi} consistently show that the tails of CGCG 97087N and the gas of UGC 6697 to which they connect is
shocked (enhanced [NII]$\lambda6684$/H$\alpha$ and [OI]$\lambda6300$/H$\alpha$). In the tail of UGC 6697, the gas 
is characterized by the lowest values of [NII]$\lambda6684$/H$\alpha$, possibly because of its low metallicity, but the high [OI]$\lambda6300$/H$\alpha$ suggests
that also this gas is shocked and turbulent. Hence, all stripped gas shows some evidence of shocks  which contribute to the ionization.
The fact that the strongest indications are found in the tails of CGCG 97087N and in the gas connecting the two galaxies hints
at a physical interaction. Similar conclusions were drawn also from the distribution of the velocity dispersion in this regions
both in the single and in the double velocity component fit.\\
\section{The HII regions}
\label{sec_hii}
\begin{figure*}
\centering
\includegraphics[width=18cm]{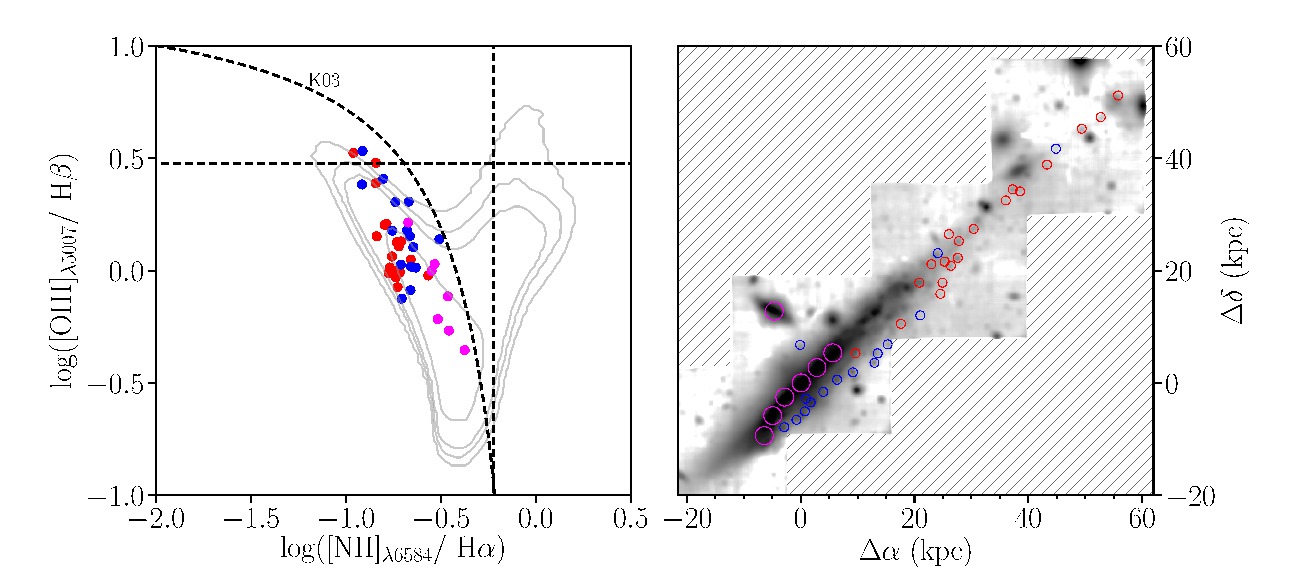}
\caption{Left: BPT diagram of HII regions selected by visual inspection to the [OIII]$\lambda5007$ map. Right: the collapsed median cube smoothed 10x10 pixels
with circles indicating the position of the selected HII regions. HII regions that are part of the main body of galaxies are circled in magenta,
while HII regions outside the bright disk of the galaxy are circled in blue ($v-v_{sys}<100~\rm km~s^{-1}$) and red ($v-v_{sys}>100~\rm km~s^{-1}$).
Note that, by such division, HII regions that lie outside the galaxy are neatly distributed along two separate trails.
}      
\label{hii}  
\end{figure*}
Compact knots of star formations lie outside the stellar disk
along the northern and, to a lesser extent, southern periphery of the tail. Gas in and around these HII region is 
characterized by lower velocity dispersions than the surrounding medium. It is therefore interesting to study in more details the properties of these compact knots.
First, by visual inspecting the map of the [OIII]$\lambda5007$ (the faintest line of the BPT), we select the compact knots and extract 
line fluxes in circular apertures, along with the average velocity and velocity dispersion 
of the H$\alpha$ line. 
In order to test whether these compact knots are associated also to continuum emission, we collapse the  MUSE cube on the 
lambda dimension and search for counterpats by visual inspection.\\ 
Moreover, we cover the brightest regions of the main body of UGC 6697 and of CGCG 97087N with 7 
circular apertures in order to characterize the properties of the star forming regions found inside the galaxy with respect to the extra-disk knots.
We measure the metallicity from the O3N2 ratio 
with the relation calibrated by \citet{curti16}.
The extinction is computed from the theoretical value of the $\rm H\alpha/H\beta$ ratio 
considering the selective extinction of H$\alpha$ relative to H$\beta$ as f(H$_\alpha$) = -0.297 evaluated from the Galactic extinction law of \citet{cardelli}.
In Fig. \ref{hii}, we show the BPT diagram of the HII regions identified by visual inspection.
All regions appear to be normal star forming clumps in the diagram, and we do not detect any remarkable difference between the HII regions in the main body of the 
galaxies and the external ones.

Furthermore, no remarkable differences between the extra-disk HII regions and the galactic HII regions are found in metallicity and extinction.
The gas ionized inside the disk has similar metallicities compared to the ones outside the galaxy.
In the disk, the average metallicity  ($\rm <12+log(O/H)>\approx8.6$) is close to solar, similarly to the HII regions outside the galaxy with metallicity 
between 8.4 and 8.6, in agreement with the values extracted by \citet{pg01}. 
Extinction is similarly consistent in both the onboard and stripped HII regions, and it is almost negligible ($\rm <A_\nu>=0.4 \rm mag$).

In the velocity maps of H$\alpha$, most of the HII regions in the back tail are found at velocities larger than $100 ~\rm km~s^{-1}$.
If we separate all HII regions with such velocities from the remaining ones we obtain the separation of red and blue symbols in Fig.\ref{hii}.
Looking separately at the distribution of compact knots circled in blue and red, we note that these two are neatly 
positioned along two parallel paths running at the northern and southern boundary of the trailing gas. 
This suggests that there are two preferential streams along which star formation can occur in the stripped gas, which
we observe in superposition due to projection effects. 
This evidence hints at a physical mechanism able to funnel along preferential tracks the gas. 
Such an effect has been proposed by \citet{dursi08} as due to cluster magnetic field draping.
\section{Discussion and conclusions}
\label{sec_discuss}
In this section we present a comprehensive picture of the turbulent life of UGC 6697 in the cluster 
that we discuss in the context of previous environmental studies.
\subsection{Which process is driving the stripping?}
As discussed in section 2, UGC 6697 has been under investigation starting from the early works of Gavazzi (1978), till 
the recent deep data of SUBARU \citep{yagi17} and present MUSE observations. Throughout the years, hydrodynamic interactions between the ISM ad the ICM have been 
invoked as the main culprit of the asymmetries in the young stellar distribution \citep[i.e. in jelly-fish galaxies, ][]{poggianti16} and in the gaseous distribution of UGC 6697.\\
Mainly because of the complex kinematics in the gaseous component of the galaxy, Gavazzi et al. (2001) proposed a different scenario in which UGC 6697 
is a system of two merging galaxies. We tend to exclude this possibility since we can ascribe the complex gas kinematics 
to the superposition of emitting gas with different velocities along the line of sight while being stripped from the SE and dragged to NW (see Fig. \ref{cartoon}). 
In the maps of the stellar kinematics of UGC 6697 we do not find evidences of two kinematically separated systems.
Hence we suggest that UGC 6697 is a single massive (v$_{rot}\sim 250\rm km~s^{-1}$) edge-on spiral galaxy suffering from the hydrodynamic interaction with the ICM.\\

Taking advantage of the exquisite MUSE observations we have indeed measured and separated the gas emission of the two different
components observed along the line of sight.
Comparing the velocity of gas and stars we separate the ionized gas in a stripped component and a component still bound to the galaxy.
The gas still onboard the galaxy has low velocity dispersions ($\le 50\rm ~km~s^{-1}$) while
the stripped gas is characterized by higher velocity dispersions ($60-100\rm ~km~s^{-1}$) with the exception of some locations harboring HII regions. 
This  indicates that the stripped gas is characterized by turbulent motions and shocks, the presence of which is also indicated by
the high values of [OI]/H$\alpha$ ratio.
High velocity dispersions an enhanced [OI]/H$\alpha$ ratio in the stripped gas are also found in other galaxies suffering ram pressure stripping 
such as ESO 137-001 \citep{fuma14} or the galaxies of the Shapley supercluster described in \citet{merluzzi16}.\\
Owing to our kinematical decomposition we 
evaluated from the H$\alpha$ luminosity of the stripped component that the interaction efficiently stripped  $\approx 10^{9} M_\odot$ of ionized gas, 
consistent with the mass loss of $\approx 10^{9} M_\odot$  implied by the  HI deficiency parameter ($Def_{HI}=0.23$; Giovanelli \& Haynes 1985) measured by \citet{pg89}.
Therefore, despite the high mass and the edge-on configuration of UGC 6697, the stripping remains efficient.\\

The lower mass companion galaxy of UGC 6697, CGCG 97087N, is very close in projection to UGC 6697 ($\rm \approx 14 kpc$) and
only the recent deep SUBARU observations \citep{yagi17} unveiled a double tail trailing behind the galaxy and connecting to the disk of UGC 6697.
Owing to the sensitivity of MUSE we clearly detect both tails that connect to the gas of UGC 6697 but in the velocity space they remain well separated.
The tails indicate that also CGCG 97087N is suffering ram pressure from the ICM of the cluster while transiting in the cluster, similarly to UGC 6697.
Also CGCG 97087N is detected in X-ray (although no tails are observed, see Fig. \ref{xrayim}) by \citet{sun05} who speculate that the X-ray emission arises from the gas 
heated by active star formation triggered by the tidal force of UGC 6697.
The large difference between the recessional velocity of UGC 6697 and CGCG 97087N ($\Delta(cz)\approx 800 \rm~ km~s^{-1}$ ) suggests that, if any, 
only a mild gravitational interaction occurred, consistent with the absence of 
stellar streams connecting the two galaxies along the gaseous tails even in the deepest data available of the SUBARU. 
However our non-parametrical analysis of the H$\alpha$ flux reveals that the tails of CGCG 97087N 
penetrate deeply in the signal of the main body of UGC 6697. 
Moreover, in the [NII]/H$\alpha$ map the tails of 
of CGCG 97087N have enhanced values of the ratio and connect perfectly to the only region of the stripped gas of UGC 6697 with such [NII]/H$\alpha$ value.
At the same time, the velocity dispersion of the tails appears connected to that of UGC 6697.
Such indicators are consistent with a past galaxy-galaxy interaction that shocked the gaseous component of both galaxies (enhanced [NII]/H$\alpha$)  while
leaving the stellar distributions unaltered. The impact between the galaxies occurred on almost perpendicular orbits, as hinted by the directions of the tails. Hence,
the hydrodynamic interaction must have been short and the ram pressure on CGCG 97087N only mildly enhanced for a very short time.
We further speculate that such interaction possibly helped in keeping the stripping process of UGC 6697 efficient by perturbing its potential well.\\

Other gravitational interactions, i.e. with the giant elliptical NGC 3842, cannot be excluded  $a~priori$.
Indeed, the deep SUBARU data (see Fig. \ref{subaru}) highlight the proximity of one of the two giant E galaxies of Abell 1367, NGC 3842 whose recessional velocity
differs from UGC 6697 by only $\approx 500\rm~ km~s^{-1}$. This galaxy lies at a projected distance of $\approx98 kpc$ from the center of UGC 6697. 
The tail of UGC 6697 suggests a diagonal motion from NW to SE, while the bended jets of NGC 3842 (a Narrow Angle Tail, NAT, galaxy) hints at a W-E motion.
The small difference in the recessional velocities of the two galaxies and their small projected distance suggests a long interaction time.
However, the real-space relative distances, relative velocities, and orbits are unknown preventing a reliable determination of parameters such as 
the time of encounter or the truncation radius, that would quantify the degree of the tidal perturbation.
Given the mild asymmetries in the NW back of UGC 6697, we cannot completely exclude that a contribution from the tidal field of NGC 3842 
is helping the stripping process. However the absence of clear stellar streams, tidal tails, and shells in and around the galaxy disfavours the idea that gravitational interactions 
are playing a major role in this process.
\subsection{The tail(s)}
\begin{figure}
\centering
\includegraphics[width=9cm]{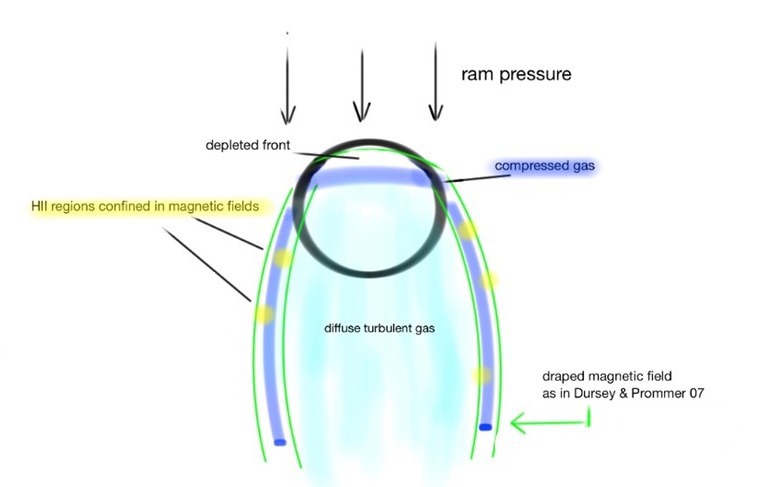}
\caption{Cartoon of the derived toy-model for UGC 6697. }      
\label{modelhii}  
\end{figure}
In our observations, the properties of the tails belonging to UGC 6697 and CGCG 97087N  are different. 
In particular, the tail of UGC 6697 harbors many HII regions neatly distributed along preferential trails while the tail of CGCG 97087N does not. 
Looking at the kinematics of the two tails, the striking difference between the two galaxies is the presence  of extended regions of the gas 
that retains small velocity dispersions ($\rm <50~km~s^{-1}$) in the tail of UGC 6697. The typical velocity dispersions in the double tails of CGCG 97087N
are of the order $80-100 \rm ~km~s^{-1}$ while in UGC 6697 the gas associated to the trailing HII regions has typical velocity dispersions below $50 \rm ~km~s^{-1}$.
At the same time, in the tails of CGCG 97087N the [OI]/H$\alpha$ and [NII]/H$\alpha$ are highly enhanced with respect to the ionized gas in the whole frame, 
suggesting the presence of shocked gas and turbulent motions.\\
The gas stripped from UGC 6697 also shows regions where the [OI]/H$\alpha$ ratio suggests that shocks are heating the gas, however the [NII]/H$\alpha$ ratio
of the stripped gas is on average low with the exception of the region of possible interaction with CGCG 97097N. 
Taking into account the models at different metallicities of \citet{kew01} we interpret the low values of [NII]/H$\alpha$ as lower 
metallicity of the stripped diffuse gas.
This is consistent with a picture where the stripping acted efficiently  on the most external gas of the 
galaxy which is characterized by lower metallicities ($\approx 0.4 Z_\odot$), consistent with the metallicity gradients observed in spiral Sb-Sc galaxies \citep{zcalifa}.\\
Our results are in line with the one observed in the  ESO 137-001 \citep{fuma14,fox16} galaxy where the  stripped tail shows enhanced [OI]/H$\alpha$ and [NII]/H$\alpha$ as well as high velocity dispersions 
with the exception of zones harboring HII regions. 
Consistent results have also recently been found within two galaxies showing tails of ram pressure stripping within the  GASP survey  \citep{gasp1,gasp2}.
Hence, the star formation in the tail appears to be associated with the less-turbulent component of the stripped gas.
The properties (i.e., line ratios and velocity dispersions) of the HII regions that formed outside the galaxy do not differentiate from the HII regions in the galaxy.
Apparently, these are formed only along two preferential trails suggesting that in these regions cooling and collapse are favored possibly by 
an enhanced gas density and magnetic confinement. 
In Fig. \ref{modelhii} (left panel) we plot a cartoon of our toy-model of the UGC 6697 system that we developed in the light 
of recent magneto hydrodynamical simulations by \citet{magneto}.\\
In this model, UGC 6697 is traveling at high speed near the core of Abell 1367 in a nearly edge-on configuration.
This configuration should disfavour ram pressure stripping  but, on the contrary,
the stripping may be enhanced and favored by the interaction with CGCG 97087N and by the interaction of the disk surface by the magnetic fields sliding past the ISM/ICM interface \citep{magneto}.
This configuration of impact is similar to the ones explored by the simulations by \citet{magneto} of a $\sim10^{11}M_\odot$ stellar mass galaxy suffering from magneto-hydrodynamic
interaction with an ICM with physical parameters matching those of observed clusters ($n_e\sim 0.5\times10^{-3}\rm cm^{-3}$; T$\sim 7\rm keV$; B$\sim 1.7 \rm \mu G$) at a distance of $\approx 0.6$ Mpc from the center of the cluster.
These simulations show that the presence of magnetic fields induces a filamentary shape of the stripped material that would otherwise appear as clumpy. 
In particular, two magnetized tails with enhanced gas density arise in the simulations, favoring cooling and collapse with respect to the turbulent
and less dense gas in the other filaments. 
We speculate that in UGC 6697 we are looking at a very similar phenomenon where the motion of the galaxy through the magnetized ICM drapes the magnetic fields lines \citep{dursi08} and
produces two magnetized trails similar to the ones in the simulations of \citet{magneto}. At the front of the galaxy, ram pressure compresses the gas in the NW direction and leaves a depleted front.
At the same time, along the body of the galaxy, at the interface between the ISM and the ICM, Kelvin-Helmotz instabilities and magnetic scraping rip the gas from the disk. 
Subsequently, the wind of ram pressure pushes this gas to the back of the galaxy. In the magnetized tails, the gas is compressed, cools and forms stars.\\
\begin{acknowledgements}
The authors would like to thank the referee for his constructive criticism.
We also thank M. Sun who kindly shared his X-ray data with us.
This research has made use of the GOLDmine database (Gavazzi et
al. 2003, 2014b) and of the NASA/IPAC Extragalactic Database (NED) which is
operated by the Jet Propulsion Laboratory, California Institute of Technology, under contract with the National Aeronautics and Space Administration.  
M. Fossati acknowledges the support of the Deutsche Forschungsgemeinschaft via Projects WI 3871/1-1, and WI 3871/1-2.
M. Fumagalli acknowledges support by the Science and Technology Facilities Council [grant number ST/P000541/1].
For access to the codes and advanced data products used in this work, please contact the authors or visit \url{https://github.com/mifumagalli/mypython/tree/master/ifu}.
Raw data are publicly available via the ESO Science Archive Facility.
\end{acknowledgements}


\begin{thebibliography}{}
\bibitem[Allen et al.(2008)]{allen08} Allen, M.~G., Groves, B.~A., Dopita, M.~A., Sutherland, R.~S., \& Kewley, L.~J.\ 2008, \apjs, 178, 20-55
\bibitem[Baldwin et al.(1981)]{bpt} Baldwin, J.~A., Phillips, M.~M., \& Terlevich, R.\ 1981, \pasp, 93, 5 
\bibitem[Bekki \& Couch(2003)]{bekki03} Bekki, K., \& Couch, W.~J.\ 2003, \apjl, 596, L13
\bibitem[Binggeli et al.(1987)]{bing87} Binggeli, B., Tammann, G.~A., \& Sandage, A.\ 1987, \aj, 94, 251 
\bibitem[Binney \& Tremaine(1987)]{binney_tremaine} Binney, J., \& Tremaine, S.\ 1987, Princeton, NJ, Princeton University Press, 1987, 747 p.
\bibitem[Bellhouse et al.(2017)]{gasp2} Bellhouse, C., Jaffe, Y.~L., Hau, G.~K.~T., et al.\ 2017, arXiv:1704.05087
\bibitem[Bliton et al.(1998)]{nat} Bliton, M., Rizza, E., Burns, J.~O., Owen, F.~N., \& Ledlow, M.~J.\ 1998, \mnras, 301, 609
\bibitem[Boselli et al.(1994)]{1994A&A...285...69B} Boselli, A., Gavazzi, G., Combes, F., Lequeux, J., \& Casoli, F.\ 1994, \aap, 285,69  
\bibitem[Boselli \& Gavazzi(2006)]{bos06} Boselli, A., \& Gavazzi, G.\ 2006, \pasp, 118, 517 
\bibitem[Boselli et al.(2008)]{bos08} Boselli, A., Boissier, S., Cortese, L., \& Gavazzi, G.\ 2008, \apj, 674, 742-767
\bibitem[Boselli \& Gavazzi(2014)]{bos14} Boselli, A., \& Gavazzi, G.\ 2014, \aapr, 22, 74 
\bibitem[Boselli et al.(2016)]{1690} Boselli, A., Cuillandre, J.~C., Fossati, M., et al.\ 2016, \aap, 587, A68 
\bibitem[Cappellari \& Emsellem(2004)]{ppxf} Cappellari, M., \& Emsellem, E.\ 2004, \pasp, 116, 138 
\bibitem[Cardelli et al.(1989)]{cardelli} Cardelli, J.~A., Clayton, G.~C., \& Mathis, J.~S.\ 1989, \apj, 345, 245
\bibitem[Chabrier(2003)]{chabrier} Chabrier, G.\ 2003, \pasp, 115, 763
\bibitem[Chung et al.(2009)]{VIVA} Chung, A., van Gorkom, J.~H., Kenney, J.~D.~P., Crowl, H., \& Vollmer, B.\ 2009, \aj, 138, 1741 
\bibitem[Consolandi et al.(2016)]{C16} Consolandi, G., Gavazzi, G., Fumagalli, M., Dotti, M., \& Fossati, M.\ 2016, \aap, 591, A38
\bibitem[Courteau et al.(2007)]{courteau07} Courteau, S., Dutton, A.~A., van den Bosch, F.~C., et al.\ 2007, \apj, 671, 203 
\bibitem[Cowie \& McKee(1977)]{cowie77} Cowie, L.~L., \& McKee, C.~F.\ 1977, \apj, 211, 135 
\bibitem[Curti et al.(2017)]{curti16} Curti, M., Cresci, G., Mannucci, F., et al.\ 2017, \mnras, 465, 1384
\bibitem[Draine(2011)]{ism} Draine, B.~T.\ 2011, Physics of the Interstellar and Intergalactic Medium by Bruce T.~Draine.~Princeton University Press, 2011.~ISBN: 978-0-691-12214-4
\bibitem[Dressler(1980)]{dress80} Dressler, A.\ 1980, \apj, 236, 351 
\bibitem[Duc et al.(2015)]{duc15} Duc, P.-A., Cuillandre, J.-C., Karabal, E., et al.\ 2015, \mnras, 446, 120 
\bibitem[Dursi \& Pfrommer(2008)]{dursi08} Dursi, L.~J., \& Pfrommer, C.\ 2008, \apj, 677, 993-1018
\bibitem[Ebeling et al.(2014)]{ebeling14} Ebeling, H., Stephenson, L.~N., \& Edge, A.~C.\ 2014, \apjl, 781, L40
\bibitem[Fontanot et al.(2009)]{fonta09} Fontanot, F., De Lucia, G., Monaco, P., Somerville, R.~S., \& Santini, P.\ 2009, \mnras, 397, 1776
\bibitem[Fossati et al.(2012)]{fox12} Fossati, M., Gavazzi, G., Boselli, A., \& Fumagalli, M.\ 2012, \aap, 544, A128
\bibitem[Fossati et al.(2016)]{fox16} Fossati, M., Fumagalli, M., Boselli, A., et al.\ 2016, \mnras, 455, 2028 
\bibitem[Fritz et al.(2017)]{gasp3} Fritz, J., Moretti, A., Poggianti, B., et al.\ 2017, arXiv:1704.05088
\bibitem[Fruscione \& Gavazzi(1990)]{1990BAAS...22..746F} Fruscione, A., \& Gavazzi, G.\ 1990, \baas, 22, 746 
\bibitem[Fumagalli et al.(2011)]{fumama12} Fumagalli, M., Gavazzi, G., Scaramella, R., \& Franzetti, P.\ 2011, \aap, 528, A46
\bibitem[Fumagalli et al.(2014)]{fuma14} Fumagalli, M., Fossati, M., Hau, G.~K.~T., et al.\ 2014, \mnras, 445, 4335 
\bibitem[Fumagalli et al.(2017)]{fuma17} Fumagalli, M., Haardt, F., Theuns, T., et al.\ 2017, \mnras, 467, 4802 
\bibitem[Gavazzi(1978)]{1978A&A....69..355G} Gavazzi, G.\ 1978, \aap, 69, 355 
\bibitem[Gavazzi et al.(1984)]{1984A&A...137..235G} Gavazzi, G., Tarenghi, M., Jaffe, W., Boksenberg, A., \& Butcher, H.\ 1984, \aap, 137, 235 
\bibitem[Gavazzi \& Jaffe(1987)]{1987A&A...186L...1G} Gavazzi, G., \& Jaffe, W.\ 1987, \aap, 186, L1 
\bibitem[Gavazzi(1987)]{1987ApJ...320...96G} Gavazzi, G.\ 1987, \apj, 320, 96 
\bibitem[Gavazzi(1989)]{pg89} Gavazzi, G.\ 1989, \apj, 346, 59 
\bibitem[Gavazzi et al.(1995)]{1995A&A...304..325G} Gavazzi, G., Contursi, A., Carrasco, L., et al.\ 1995, \aap, 304, 325 
\bibitem[Gavazzi et al.(2001)]{pg01} Gavazzi, G., Marcelin, M., Boselli, A., et al.\ 2001, \aap, 377, 745 
\bibitem[Gavazzi et al.(2003)]{2003A&A...400..451G} Gavazzi, G., Boselli, A., Donati, A., Franzetti, P., \& Scodeggio, M.\ 2003, \aap, 400, 451 
\bibitem[Gavazzi(2009)]{pg09} Gavazzi, G.\ 2009, Revista Mexicana de Astronomia y Astrofisica Conference Series, 37, 72
\bibitem[Gavazzi et al.(2013)]{pg13} Gavazzi, G., Savorgnan, G., Fossati, M., et al.\ 2013, \aap, 553, A90
\bibitem[Gavazzi et al.(2014b)]{2014arXiv1401.8123G} Gavazzi, G., Franzetti, P., \& Boselli, A.\ 2014b, arXiv:1401.8123 
\bibitem[Giovanelli \& Haynes(1985)]{1985ApJ...292..404G} Giovanelli, R., \& Haynes, M.~P.\ 1985, \apj, 292, 404
\bibitem[Gonz{\'a}lez Delgado et al.(2015)]{zcalifa} Gonz{\'a}lez Delgado, R.~M., Garc{\'{\i}}a-Benito, R., P{\'e}rez, E., et al.\ 2015, \aap, 581, A103 
\bibitem[Gunn \& Gott(1972)]{1972ApJ...176....1G} Gunn, J.~E., \& Gott, J.~R., III 1972, \apj, 176, 1 
\bibitem[J{\'a}chym et al.(2014)]{jachy14} J{\'a}chym, P., Combes, F., Cortese, L., Sun, M., \& Kenney, J.~D.~P.\ 2014, \apj, 792, 11
\bibitem[Kauffmann et al.(2003)]{kauff03} Kauffmann, G., Heckman, T.~M., Tremonti, C., et al.\ 2003, \mnras, 346, 1055
\bibitem[Kenney et al.(2014)]{kenney14} Kenney, J.~D.~P., Geha, M., J{\'a}chym, P., et al.\ 2014, \apj, 780, 119
\bibitem[Kewley et al.(2001)]{kew01} Kewley, L.~J., Dopita, M.~A., Sutherland, R.~S., Heisler, C.~A., \& Trevena, J.\ 2001, \apj, 556, 121
\bibitem[Kregel et al.(2005)]{kregel05} Kregel, M., van der Kruit, P.~C., \& Freeman, K.~C.\ 2005, \mnras, 358, 503 
\bibitem[Larson et al.(1980)]{larson80} Larson, R.~B., Tinsley, B.~M., \& Caldwell, C.~N.\ 1980, \apj, 237, 692 
\bibitem[Merluzzi et al.(2016)]{merluzzi16} Merluzzi, P., Busarello, G., Dopita, M.~A., et al.\ 2016, \mnras, 460, 3345 
\bibitem[Moore et al.(1996)]{harassment} Moore, B., Katz, N., Lake, G., Dressler, A., \& Oemler, A.\ 1996, \nat, 379, 613 
\bibitem[Nilson(1973)]{1973ugcg.book.....N} Nilson, P.\ 973,    Acta Universitatis Upsaliensis.~Nova Acta Regiae Societatis Scientiarum Upsaliensis - Uppsala Astronomiska Observatoriums Annaler, Uppsala: Astronomiska Observatorium, 1973,  
\bibitem[Noll et al.(2014)]{2014A&A...567A..25N} Noll, S., Kausch, W., Kimeswenger, S., et al.\ 2014, \aap, 567, A25
\bibitem[Nulsen(1982)]{Nulsen82} Nulsen, P.~E.~J.\ 1982, \mnras, 198, 1007
\bibitem[Osterbrock(1989)]{oster89} Osterbrock, D.~E.\ 1989, Research supported by the University of California, John Simon Guggenheim Memorial Foundation, University of Minnesota, et al.~Mill Valley, CA, University Science Books, 1989, 422 p.
\bibitem[Osterbrock \& Ferland(2006)]{oster06} Osterbrock, D.~E., \& Ferland, G.~J.\ 2006, Astrophysics of gaseous nebulae and active galactic nuclei, 2nd.~ed.~by D.E.~Osterbrock and G.J.~Ferland.~Sausalito, CA: University Science Books
\bibitem[Piffaretti et al.(2011)]{piff11} Piffaretti, R., Arnaud, M., Pratt, G.~W., Pointecouteau, E., \& Melin, J.-B.\ 2011, \aap, 534, A109 
\bibitem[Poggianti et al.(2016)]{poggianti16} Poggianti, B.~M., Fasano, G., Omizzolo, A., et al.\ 2016, \aj, 151, 78
\bibitem[Poggianti et al.(2017)]{gasp1} Poggianti, B.~M., Moretti, A., Gullieuszik, M., et al.\ 2017, arXiv:1704.05086 
\bibitem[Read et al.(2006)]{read06} Read, J.~I., Wilkinson, M.~I., Evans, N.~W., Gilmore, G., \& Kleyna, J.~T.\ 2006, \mnras, 366, 429
\bibitem[Rich et al.(2011)]{rich11} Rich, J.~A., Kewley, L.~J., \& Dopita, M.~A.\ 2011, \apj, 734, 87
\bibitem[Ruszkowski et al.(2014)]{magneto} Ruszkowski, M., Br{\"u}ggen, M., Lee, D., \& Shin, M.-S.\ 2014, \apj, 784, 75
\bibitem[Salpeter(1955)]{salpeter} Salpeter, E.~E.\ 1955, \apj, 121, 161
\bibitem[Sarzi et al.(2006)]{gandalf} Sarzi, M., Falc{\'o}n-Barroso, J., Davies, R.~L., et al.\ 2006, \mnras, 366, 1151 
\bibitem[Scott et al.(2013)]{2013MNRAS.429..221S} Scott, T.~C., Usero, A., Brinks, E., et al.\ 2013, \mnras, 429, 221 
\bibitem[Siudek et al.(2017)]{siu16} Siudek, M., Ma{\l}ek, K., Scodeggio, M., et al.\ 2017, \aap, 597, A107
\bibitem[Storey \& Zeippen(2000)]{lineratios} Storey, P.~J., \& Zeippen, C.~J.\ 2000, \mnras, 312, 813 
\bibitem[Sullivan et al.(1981)]{sulli81} Sullivan, W.~T., III, Bates, B., Bothun, G.~D., \& Schommer, R.~A.\ 1981, \aj, 86, 919
\bibitem[Sun \& Vikhlinin(2005)]{sun05} Sun, M., \& Vikhlinin, A.\ 2005, \apj, 621, 718 
\bibitem[Sun et al.(2007)]{sun07} Sun, M., Donahue, M., \& Voit, G.~M.\ 2007, \apj, 671, 190
\bibitem[Yagi et al.(2007)]{yagi07} Yagi, M., Komiyama, Y., Yoshida, M., et al.\ 2007, \apj, 660, 1209
\bibitem[Yagi et al.(2010)]{yagi10} Yagi, M., Yoshida, M., Komiyama, Y., et al.\ 2010, \aj, 140, 1814
\bibitem[Yagi et al.(2017)]{yagi17} Yagi, M.,Yoshida, M., Gavazzi, G.,  et al.\ 2017, submitted to \apj
\bibitem[Yoshida et al.(2004)]{yoshi04} Yoshida, M., Ohyama, Y., Iye, M., et al.\ 2004, \aj, 127, 90 
\bibitem[Vazdekis et al.(2010)]{miles_lib} Vazdekis, A., S{\'a}nchez-Bl{\'a}zquez, P., Falc{\'o}n-Barroso, J., et al.\ 2010, \mnras, 404, 1639
\bibitem[Wisnioski et al.(2015)]{wisnioski15} Wisnioski, E., F{\"o}rster Schreiber, N.~M., Wuyts, S., et al.\ 2015, \apj, 799, 209 
\bibitem[Yagi et al.(2007)]{yagi07} Yagi, M., Komiyama, Y., Yoshida, M., et al.\ 2007, \apj, 660, 1209 
\bibitem[Zibetti et al.(2009)]{zibi09} Zibetti, S., Charlot, S., \& Rix, H.-W.\ 2009, \mnras, 400, 1181
\bibitem[Zwicky et al.(1961-68)]{1961cgcg.book.....Z} Zwicky, F., Herzog, E., Wild, P., Karpowicz, M., \& Kowal, C.~T.\ 1961-68, Pasadena: California Institute of Technology (CIT). 

\end{thebibliography}
\end{document}